\documentclass[prd,aps,prl,twocolumn,superscriptaddress]{revtex4-2}
\usepackage{graphicx}
\usepackage{braket}
\usepackage{amsmath}
\usepackage{relsize}
\usepackage{titlesec}
\usepackage{svg}

\ExplSyntaxOn
\NewDocumentCommand{\longdash}{ O{2} }
 {
  --\prg_replicate:nn { #1 - 1 } { \negthinspace -- }
 }
\ExplSyntaxOff

\begin{document}


\title{Secondary Lepton Production, Propagation, and Interactions with \texttt{NuLeptonSim}}



\author{Austin Cummings}
\affiliation{Departments of Physics and Astronomy $\&$ Astrophysics, Institute for Gravitation and the Cosmos,\\
Pennsylvania State University, University Park, PA 16802, USA}
  
\author{Ryan Krebs}
\affiliation{Departments of Physics and Astronomy $\&$ Astrophysics, Institute for Gravitation and the Cosmos,\\
Pennsylvania State University, University Park, PA 16802, USA}

\author{Stephanie Wissel}
\affiliation{Departments of Physics and Astronomy $\&$ Astrophysics, Institute for Gravitation and the Cosmos,\\
Pennsylvania State University, University Park, PA 16802, USA}
  
\author{Jaime Alvarez-Mu\~{n}iz}
\affiliation{Departamento de F\'{i}sica de Part\'{i}culas \& Instituto Galego de F\'{i}sica de Altas Enerx\'{i}as,\\
University de Santiago de Compostela, 15782 Santiago de Compostela, Spain}

\author{Washington R. Carvalho Jr.}
\affiliation{Faculty of Physics, University of Warsaw, Ludwika Pasteura 5, 02-093 Warsaw, Poland}

\author{Andr\'{e}s Romero-Wolf}
\affiliation{Jet Propulsion Laboratory, California Institute of Technology,
Pasadena, CA 91109, USA}

\author{Harm Schoorlemmer}
\affiliation{Nationaal Instituut voor Kernfysica en Hoge Energie Fysica (NIKHEF), Science Park, Amsterdam, The Netherlands}
\affiliation{IMAPP, Radboud University Nijmegen, Nijmegen, The Netherlands}

\author{Enrique Zas}
\affiliation{Departamento de F\'{i}sica de Part\'{i}culas \& Instituto Galego de F\'{i}sica de Altas Enerx\'{i}as,\\
University de Santiago de Compostela, 15782 Santiago de Compostela, Spain}




\begin{abstract}
Charged current interactions of neutrinos inside the Earth can result in secondary muons and $\tau$-leptons which are detectable by several existing and planned neutrino experiments through a wide variety of event topologies. Consideration of such events can improve detector performance and provide unique signatures which help with event reconstruction. In this work, we describe \texttt{NuLeptonSim}, a propagation tool for neutrinos and charged leptons that builds on the fast \texttt{NuTauSim} framework. \texttt{NuLeptonSim} considers energy losses of charged leptons, modelled both continuously for performance or stochastically for accuracy, as well as interaction models for all flavors of neutrinos, including the Glashow resonance. We demonstrate the results from including these effects on the Earth emergence probability of various charged leptons from different flavors of primary neutrino and their corresponding energy distributions. We find that the emergence probability of muons can be higher than that of taus for energies below 100 PeV, whether from a primary muon or $\tau$ neutrino, and that the Glashow resonance contributes to a surplus of emerging leptons near the resonant energy. 
\end{abstract}


\maketitle

\section{Introduction}


Several experiments aim to detect neutrinos with energies greater than a PeV both to expand the observed astrophysical flux of neutrinos~\cite{PhysRevLett.113.101101} and discover cosmogenic neutrinos expected from propagation of the highest energy cosmic rays~\cite{BERESINSKY1969423}. Several approaches are planned that together would cover the a wide energy range (PeV scale to beyond 100 EeV) and be sensitive to all flavors of neutrinos (see \cite{ACKERMANN202255} for a recent review). Proposed detection methods include: i) detection of secondaries produced by in-ice or in-water neutrino interactions via optical Cherenkov emission \cite{Aartsen:2013jdh, margiotta2022km3net, Agostini_2020}, radio emission \cite{RNOG, ARA, ARIANNA, IceCubeGen2, ANITA, PUEO} and the radar echo technique \cite{RET} and ii) detection of extensive air showers (EAS) sourced from the decay of Earth-emergent $\tau$-leptons via optical Cherenkov emission on mountains \cite{Trinity}, high-altitude balloons \cite{SPB2}, and satellites \cite{POEMMA}, via radio emission received on and through mountains \cite{BEACON, TAROGE, GRAND}, high-altitude balloons \cite{ANITA, PUEO}, and via direct particle detection \cite{TAMBO}. In all cases, the sensitivity of these experiments to neutrinos depends not only their sensitivity to the first neutrino interaction in the Earth, but also secondary interactions induced by muons and $\tau$-leptons. \\

To evaluate a proposed experiment's sensitivity to the cosmic neutrino flux, it is necessary to accurately model neutrino propagation through the Earth, taking into account all relevant interactions and energy loss processes. A wide range of different computational models have been developed for this purpose, primarily for the Earth-skimming tau neutrino channel, including \texttt{TauRunner}\cite{Safa:2019ege}, \texttt{NuPropEarth}\cite{Garcia:2020jwr}, \texttt{NuPyProp} \cite{Garg_2023}, and \texttt{NuTauSim} \cite{Alvarez:2018, Alvarez:2019}. Each of these modeling schemes provides complimentary results and agree in large part with one another, despite often using different models for neutrino cross section and inelasticity distributions, $\tau$-lepton energy loss and decay distributions, and Earth density profile \cite{Abraham_2022}. In this work, we focus specifically on the \texttt{NuTauSim} model and recent modifications to the base framework made to widen its capabilities to address lepton propagation more fully. \\

\texttt{NuTauSim} is an open source, C\texttt{++} based Monte Carlo code which simulates the propagation of tau neutrinos through the Earth, taking into account neutrino interactions, $\tau$-lepton energy losses, and potential $\tau \rightarrow \nu_{\tau} \rightarrow \tau$ regeneration processes. Compared to other neutrino propagation codes, NuTauSim was designed primarily for speed, with the intention of being able to rapidly benchmark the effects of different parameterizations of physical parameters. In its base configuration, \texttt{NuTauSim} includes the high energy extrapolations of the neutrino cross section as given in \cite{Connolly:2011vc}. $\tau$-lepton energy losses are handled exclusively via continuous approximations (detailed in section \ref{sec:stoch}), with the extrapolation of the photonuclear energy losses given by \cite{Abramowicz:1997ms} and \cite{Armesto:2004hh}. The inelasticity of a neutrino interaction is sampled using standard results calculated from  \texttt{CTEQ5} parton distributions \cite{Lai_2000} and the decay products of the $\tau$-lepton are generated using the \texttt{PYTHIA8} Monte Carlo code \cite{Sjostrand:2019zhc}, assuming a fixed (negative) polarization for the $\tau$-lepton. The grammage profile used to propagate events is calculated using the \texttt{PREM} Earth density model \cite{DZIEWONSKI1981297} with an added outer layer with a user input depth and density to model more local topographies. The \texttt{NuTauSim} framework is built in a modular way, allowing users to quantify the observable effects of custom physical models and scenarios, such as exotic cross section models or through-mountain trajectories. \\

We have enhanced the capabilities of the \texttt{NuTauSim} framework to be more broadly applicable to evaluating experimental designs beyond those targeting only the Earth-skimming tau neutrino channel, while retaining computational efficiency, accuracy, usability, and generality. This expansion aims to address two fundamental concerns: i) additional Earth-emergent particles sourced from neutrino interactions beyond solely $\tau$-leptons originating from tau neutrinos enhance the air shower detection channel and are not currently modeled by the propagation codes highlighted above ii) in-ice (or in-water) experiments typically do not fully model neutrino propagation through the Earth in their sensitivity estimates, instead considering a neutrino survival probability as a function of grammage. Such an approximation neglects energy losses from neutral current interactions, regeneration processes, and charged leptons which can propagate long distances before potentially being observed. In the latter case, secondary events have been shown to substantially increase detection capabilities of in-ice radio experiments and provide unique event topologies that must be carefully understood \cite{secondaries}. In what follows, we detail the modifications made to \texttt{NuTauSim} to form the \texttt{NuLeptonSim} framework, namely: i) the consideration of all neutrinos and charged leptons during propagation, ii) the modeling of stochastic losses of propagating charged leptons, and iii) the expansion to capture in-ice interactions. We detail the effects of each of these upgrades and discuss their implications.\\

Due to the wide parameter space available in these simulations, for all that follows, we consider a modified PREM Earth density model with a 4~km thick ice sheet, central extrapolations of the neutrino cross sections from \cite{Connolly:2011vc}, and the charged lepton energy loss models from \cite{Abramowicz:1997ms}. Preliminary effects from modifying these parameters are presented in \cite{Alvarez:2018, Alvarez:2019}.\\




\section{Emergent particles} 
The experiments that aim to measure the diffuse flux of high energy neutrinos via the Earth-skimming detection technique aim to do so predominantly through the tau neutrino channel, measuring the EAS produced by the decay of the Earth-emergent $\tau$-lepton. For this reason, many of the existing propagation codes, including \texttt{NuTauSim}, model only the propagation of tau neutrinos and $\tau$-leptons, switching between particles types upon interaction and decay, respectively. Focusing exclusively on these interactions ignores several channels that can increase the number of observed events for a given experiment and provide a nontrivial flavor impurity. We detail here the features included in \texttt{NuLeptonSim} that aim to address this point.

\subsection{Muons from $\nu_{\mu}$} \label{sec:mu_from_mu}
\begin{figure}
\includegraphics[width=0.98\linewidth]{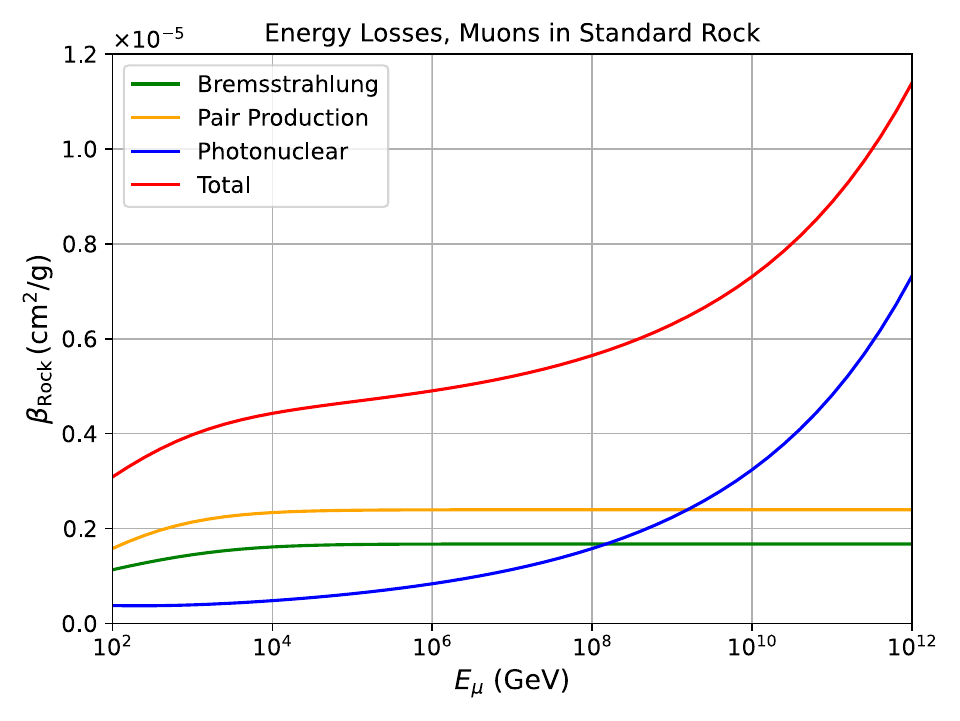}
\caption{Radiative loss factor $\beta = \frac{1}{E} \frac{dE}{dX}$ for muons in standard rock as a function of energy. Shown are the individual curves for Bremsstrahlung \cite{Petrukhin:1968}, $e^{\pm}$ pair production \cite{Kokoulin:1971}, and photonuclear \cite{Abramowicz:1997ms} interactions as well as the total energy loss factor that is used in \texttt{NuLeptonSim}.  \label{fig:muon_beta}}
\end{figure}

The impact of muon neutrinos has been shown to be non-negligible for neutrino energies below 100~PeV for both sub-orbital and orbital observation altitudes in the optical Cherenkov regime. This is due to the extended lifetime of the muon allowing for amplified Earth emergence probabilities despite increased energy losses with respect to the $\tau$-lepton \cite{Cummings:2021}. To consider propagation of muon neutrinos in the \texttt{NuLeptonSim} framework, we have updated the parameterizations of the neutrino interaction cross section to reach lower energies where muons can still substantially propagate without decaying ($\sim 100$~GeV). Below neutrino energies of 1~PeV, it is necessary to distinguish between the cross section for neutrinos and anti-neutrinos. In the \texttt{NuLeptonSim} framework, we use the results of \cite{Gandhi_1996} to model these differences. To model the energy losses of the muon, we use the continuous energy loss approximation, and calculate the total radiative loss factor $\beta$ using given parameterizations of cross section for Bremsstrahlung \cite{Petrukhin:1968}, $e^{\pm}$ pair-production \cite{Kokoulin:1971} and photonuclear \cite{Abramowicz:1997ms} interactions (see section \ref{sec:stoch}). For consistency with the results of \texttt{NuTauSim}, we use the same parameterization for the high energy extrapolation of the photonuclear cross section. The muon radiative loss factor $\beta$ used in \texttt{NuLeptonSim} is shown in Figure \ref{fig:muon_beta} as a function of muon energy.\\

\begin{figure}
\includegraphics[width=0.98\linewidth]{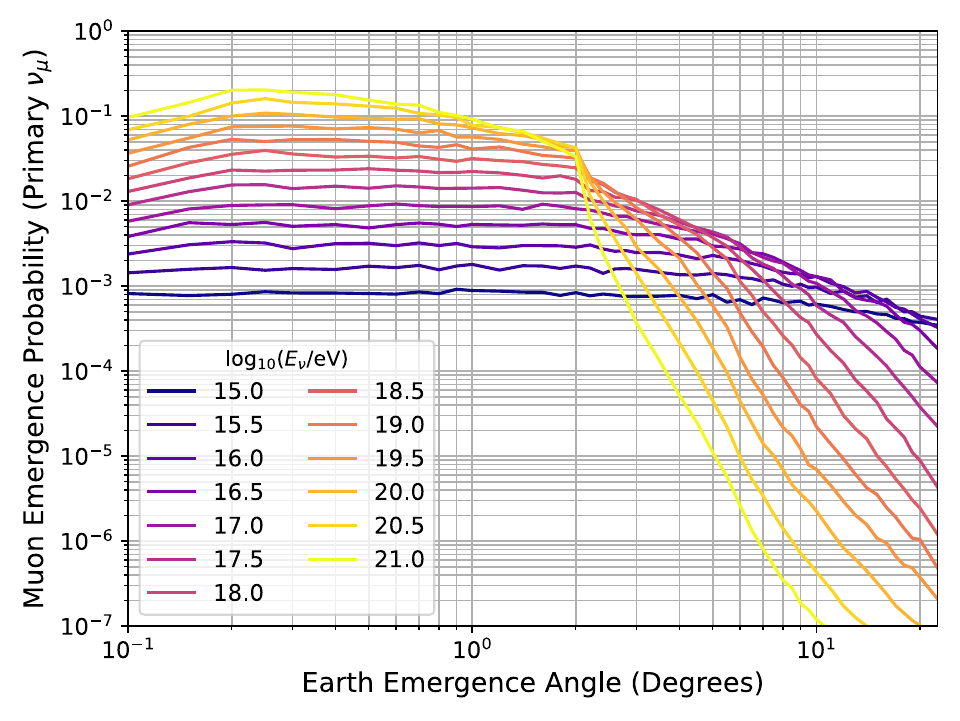}
\includegraphics[width=0.98\linewidth]{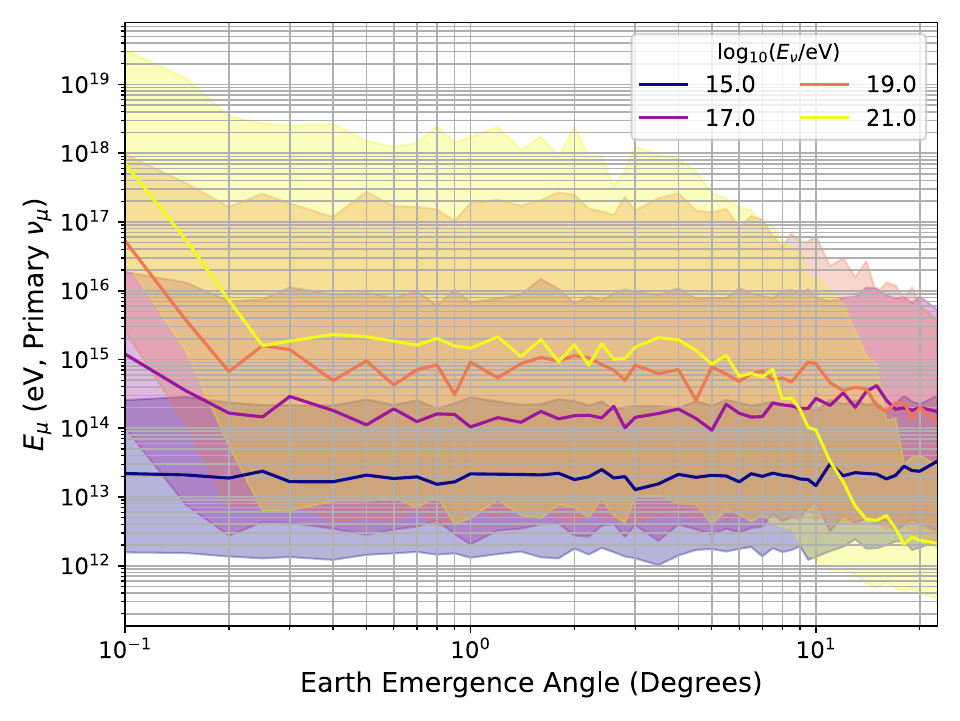}
\caption{Top: Earth emergence probability of muons with energies $E_{\mu}>10$~GeV sourced from primary mono-energetic muon neutrinos. Bottom: energy distribution of Earth-emergent muons, where the solid line represents the mean of the distribution, and the shaded regions represent the $\pm 1\sigma$ deviations. The aggregate effect on a given flux is shown in Appendix A. \label{fig:muon_emergence_probability_muons}}
\end{figure}

Lastly, for muons which propagate in \texttt{NuLeptonSim}, we sample the decay length from an exponential distribution with a mean of $\frac{E_{\mu}}{m_{\mu}} c \tau_{\mu}$, where $E_{\mu}$ is the energy of the muon, $m_{\mu}$ is the muon mass (105.6~MeV), $c$ is the speed of light, and $\tau_{\mu}$ is the average lifetime of the muon (2.2~$\mu$s). The average decay length of a muon equates to $6.25 \times 10^{8} ~ \mathrm{km} (E_{\mu}/100~\mathrm{PeV})$, which is roughly $10^{8}$ times larger than the decay length of a comparably energetic $\tau$-lepton. In Figure \ref{fig:muon_emergence_probability_muons}, we show the Earth emergence probability of muons (the ratio of Earth-emergent muons to the number of neutrinos simulated) given mono-energetic fluxes of primary muon neutrinos as well as their energy distributions as a function of Earth emergence angle, as simulated with \texttt{NuLeptonSim}.\\

The results presented in Figure \ref{fig:muon_emergence_probability_muons} agree with those of \cite{Cummings:2021}, which we summarize here. For neutrino energies below 100~PeV, the muon emergence probability is significantly amplified (nearly a factor of 600 times greater at 1~PeV) with respect to the $\tau$-lepton emergence probability due to the extended lifetimes of the muon. For neutrino energies above 100~PeV and Earth emergence angles $\bar{\theta} \leq 2^{\circ}$ (in the modified PREM model, the interface between ice and rock), the Earth emergence probabilities between muons from muon neutrinos and $\tau$-leptons from tau neutrinos are roughly equivalent, with energy losses being subdominant with respect to neutrino interactions. For high energies and steeper angles, the path length the muon must travel through the Earth increases significantly, resulting in substantial energy losses. Muons decay at characteristically smaller energies than $\tau$-leptons (roughly 100~GeV scale in rock), which results in neutrinos with low energies and correspondingly low cross sections, disallowing for $\mu \rightarrow \nu_{\mu} \rightarrow \mu$ regeneration. The bottom panel of Figure \ref{fig:muon_emergence_probability_muons} shows that Earth-emergent muons have a wide spread of energies, and despite the significant energy losses experienced during propagation inside the Earth, can retain much of the energy of their parent neutrinos, particularly for small Earth emergence angles. Such geometries maximize atmospheric grammage and allow the muon a non-negligible chance of initiating an EAS through radiative losses, rather than decay \cite{Cummings:2021}. For neutrino energies below 100~PeV, where the Earth emergence probability of muons with respect to $\tau$-leptons is amplified, muon induced EAS are typically initiated significantly higher in the atmosphere than $\tau$-leptons of corresponding energies.
\subsection{Muons from $\nu_{\tau}$} \label{sec:mu_from_tau}

Earth-emergent muons can also largely be sourced via the muonic channel of the $\tau$-lepton decay: 

\begin{equation*}
\tau^{-} \rightarrow \mu^{-} + \nu_{\tau} + \bar{\nu_{\mu}}
\end{equation*}

\noindent which occurs with a branching ratio of $\sim 17.8 \%$. While this channel has been shown to be relevant source of high-energy muons in air \cite{Cummings:2021}, it is important to quantify the behavior of Earth-emergent muons sourced from this process during propagation in the Earth. Due to its short lifetime, the $\tau$-lepton decays on the distance scale $5~\mathrm{km}(E_{\tau}/100~\mathrm{PeV})$, resulting in energetic neutrinos which may reinteract deeper in the Earth (the so-called regeneration process). For the decay listed above, it is thereby necessary to follow the produced muon, its corresponding anti-neutrino, and the tau neutrino during propagation. The mechanics of these decays are modeled using the \texttt{PYTHIA8} Monte Carlo code, and provided in convenient lookup tables for sampling during propagation.\\

\begin{figure}
\includegraphics[width=0.98\linewidth]{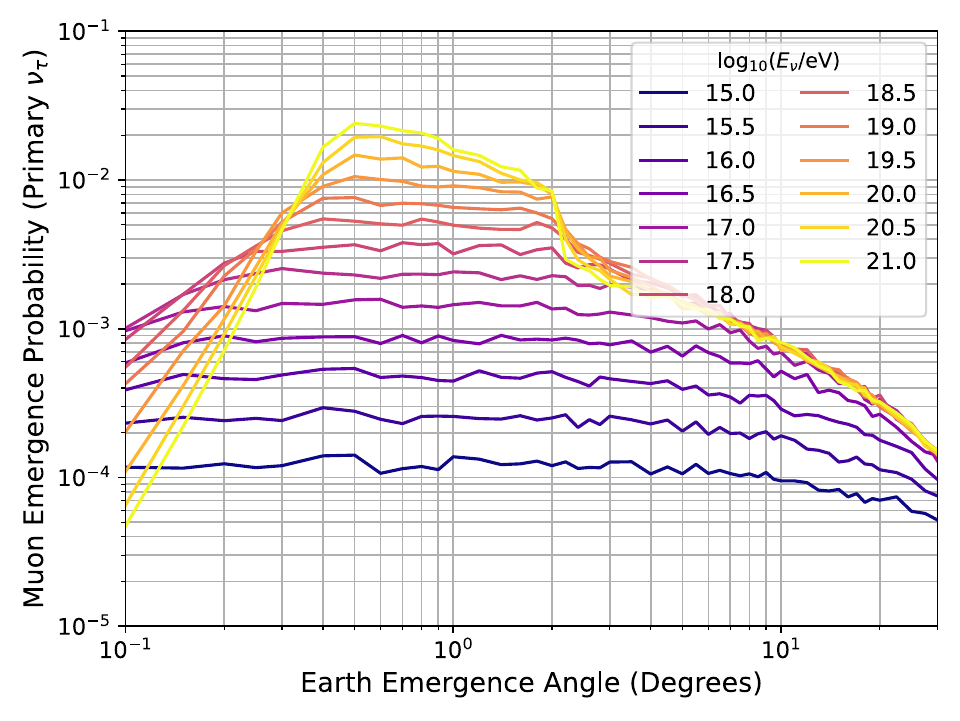}
\includegraphics[width=0.98\linewidth]{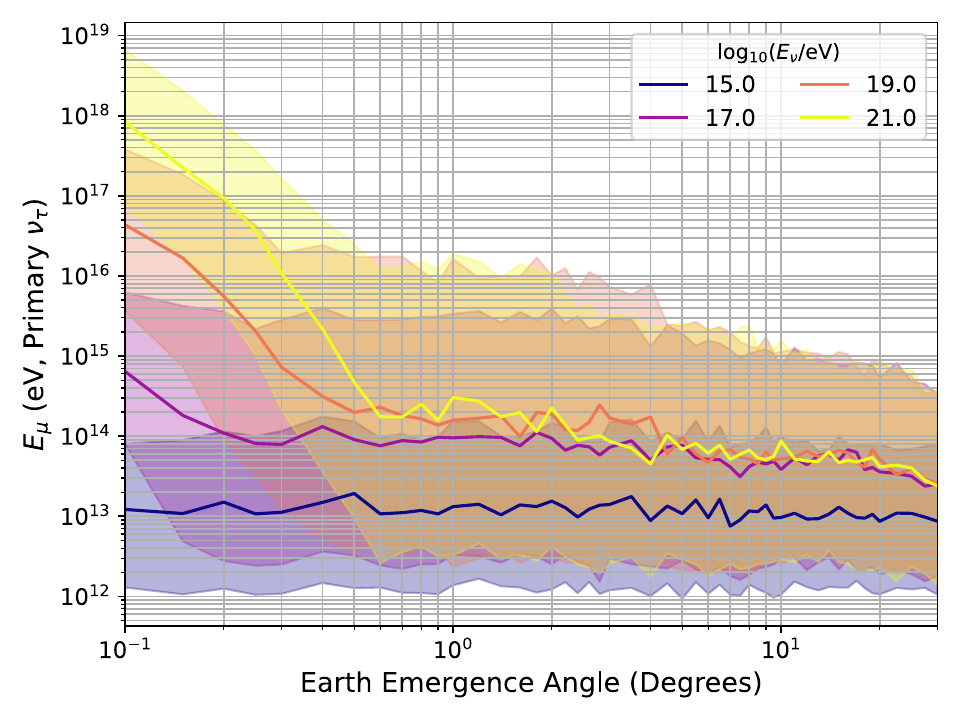}
\caption{Top: Earth emergence probability of muons with energies $E_{\mu}>10$~GeV sourced from primary mono-energetic tau neutrinos. Bottom: energy distribution of Earth-emergent muons, where the solid line represents the mean of the distribution, and the shaded regions represent the $\pm 1\sigma$ deviations. The aggregate effect on a given flux is shown in Appendix A. \label{fig:muon_emergence_probability_taus}}
\end{figure}

To track these newly created particles in \texttt{NuLeptonSim}, a ``stacking'' method was implemented into the particle propagation, whereby for any process that generates multiple particles, the interaction location, particle types, and energies are all temporarily stored in a stack. For those particles where propagation is still relevant, i.e., neutrinos (all flavors), $\tau$-leptons, and muons, they are each  propagated individually and sequentially after the parent particle has completed its propagation, either through decay, Earth emergence, or dropping below a user-defined threshold energy. In principle, this presents the possibility of multiple emergent particles from a single parent neutrino, which is included in the calculation of the Earth emergence probability. In Figure \ref{fig:muon_emergence_probability_taus}, we show the Earth emergence probability of muons given mono-energetic fluxes of primary tau neutrinos as well as their energy distributions as a function of Earth emergence angle, as simulated with \texttt{NuLeptonSim}.\\

The top panel of Figure \ref{fig:muon_emergence_probability_taus} shows that, for muon energies in the range 1-100~PeV, the emergence probability from a primary tau neutrino ranges from 50 to 2 times larger than the corresponding $\tau$-lepton emergence probability, despite the reduced probability from the decay branch of the parent $\tau$-lepton (see top panel of Figure \ref{fig:tau_comp_stoch}). This amplification also holds for higher energy primary tau neutrinos and steep Earth emergence angles ($\bar{\theta} \ge 2^{\circ}$). Both of these parameter spaces correspond to regions where $\tau$-lepton decay is likely, considering the generally reduced energies from either the parent neutrino or increased propagation ranges. For smaller Earth emergence angles, this trend reverses around primary neutrino energies of 100~PeV, where the $\tau$-lepton emergence probability begins to surpass that of the muons.\\

The mean decay length of the $\tau$-lepton increases with increasing energy and becomes comparable with the length of the trajectory through the Earth around 100~PeV in these geometries, minimizing the muon production rate. With increasing emergence angles, trajectories through the Earth are longer, allowing for exponential increases in the muon emergence probability (with exponent $2 R  \bar{\theta}/L_{\tau}$ where $R$ is the Earth radius and $L_{\tau}$ is the $\tau$-lepton decay length). The bottom panel of Figure \ref{fig:muon_emergence_probability_taus} demonstrates similar muon energy distributions to those sourced from muon neutrinos (Figure \ref{fig:muon_emergence_probability_muons}) for small Earth emergence angles, where energy losses of the $\tau$-lepton are subdominant with respect to neutrino interactions, and the decay is predominantly a probabilistic process. For Earth emergence angles $\bar{\theta} > 2^{\circ}$, the distribution in emergent muon energies becomes significantly more narrow with respect to those sourced from muon neutrinos, primarily due to the 100~PeV regeneration range for the $\tau$-lepton being the prime driver of muon production.

\subsection{Muons and $\tau$-leptons from $\bar{\nu_{e}}$ via Glashow Resonance}
A $W^{-}$ boson can be produced by the resonant interaction of an electron antineutrino with an electron--the so-called Glashow Resonance process \cite{Glashow:1960}. This cross section of this resonance peaks at electron antineutrino energies of 6.3~PeV. The average lifetime of the $W^{-}$ is extremely small, being $3 \times 10^{-25}$~s, and can reasonably be approximated to decay immediately, even at very high energies. When the $W^{-}$ boson decays, it does so to exclusively hadrons with a branching ratio of $67.7 \pm 0.3\%$ and to charged leptons and their corresponding antineutrinos with a combined branching ratio of $32.4 \pm 0.3\%$. The probability among all three leptonic flavors is nearly equivalent, giving each leptonic decay channel a branching ratio of $10.8\%$. In this way, Glashow Resonance provides another channel for the production of muons and $\tau$-leptons that should be taken into account for propagation studies, particularly for experiments sensitive to PeV-scale energies (the IceCube collaboration has recently observed an event consistent with Glashow Resonance, for example \cite{Aartsen:2021}). In addition, electron antineutrinos are produced in the decay of both muons and $\tau$-leptons, leading to potential cross-flavor regeneration processes that should be considered.\\ 

\begin{figure}
\includegraphics[width=0.98\linewidth]{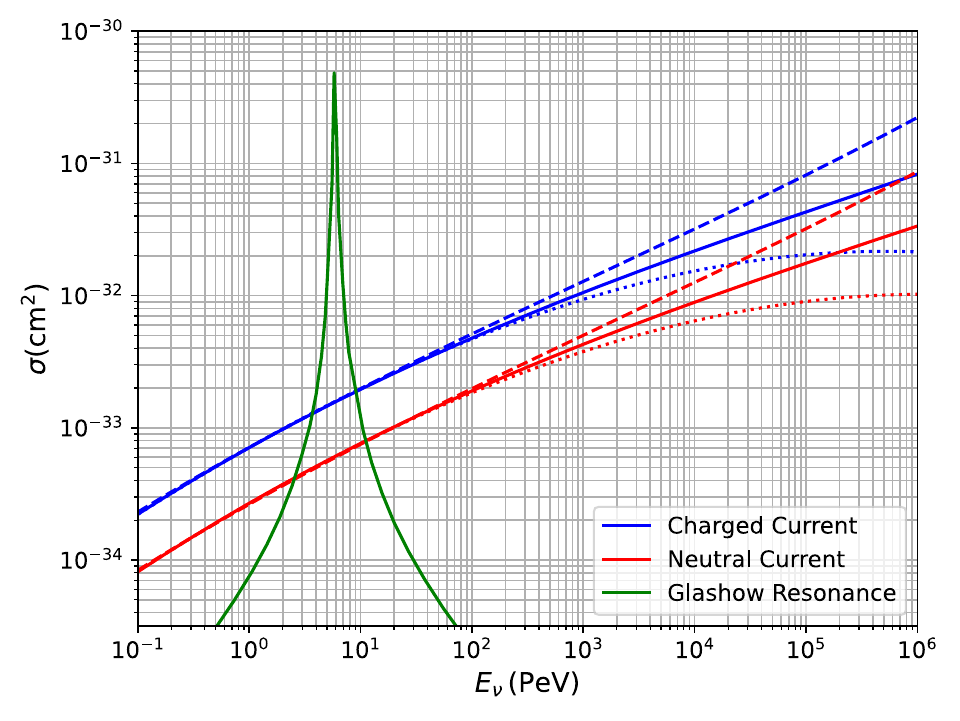}
\caption{The cross sections of the Glashow Resonance \cite{Huang_2020, PhysRevD.98.030001} and the charged and neutral current processes used in the \texttt{NuLeptonSim} framework. Dashed and doted lines show the different high energy extrapolations of the charged and neutral current cross sections included in \texttt{NuLeptonSim}. \label{fig:GR_cross_sections}}
\end{figure}

At the resonant energy of the electron antineutrino, the cross section of the Glashow Resonance process exceeds that of the corresponding charged current cross section by a factor of $\sim 400$, and quickly falls off with either increasing or decreasing energy. A comparison of the interaction cross sections for the Glashow Resonance and the charged-current and neutral-current processes is shown in Figure \ref{fig:GR_cross_sections}. Near the resonant energy, we expect nontrivial production of $W^{-}$ bosons, and therefore, muons and $\tau$-leptons, leading to potential enhancements of the Earth emergence probability within a small range of neutrino energies. More specifically, multiplying the cross section enhancement of 400 by the $10\%$ branching ratio for a given leptonic decay, we can estimate any potential enhancement to be maximally $\sim 40$ near the resonant energy.\\

The energy distribution of the charged leptons produced in the $W^{-}$ decay is well known, and depends on the helicity of the $W^{-}$. A $W^{-}$ boson produced by a high energy electron antineutrino will have a negative helicity (the spin of the $W^{-}$ is antiparallel to the momentum) \cite{Salvatore:2004}. From this, the angular distribution of the charged lepton in the decay (in the rest frame) is governed by:

\begin{equation*}
\frac{1}{N}\frac{dN}{d \mathrm{cos} \theta_{l}} = \frac{3}{16 \pi} (1+\mathrm{cos}\theta_{l})^{2}
\end{equation*}

\begin{figure}
\includegraphics[width=0.98\linewidth]{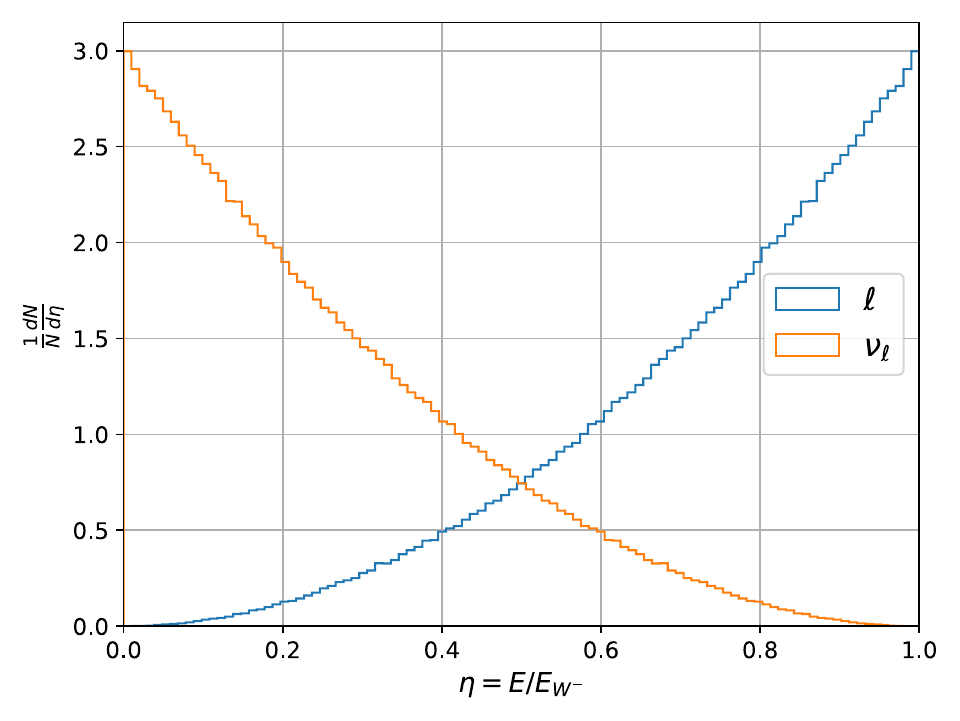}
\caption{The energy distribution of charged leptons and antineutrinos produced in the decay of a relativistic, negatively polarized $W^{-}$ used in the NuLeptonSim framework. \label{fig:W_decay_distribution}}
\end{figure}

Where $\theta_{l}$ is the angle of the charged lepton with respect to the initial momentum of the $W^{-}$. To sample the energy of a charged lepton produced in a $W^{-}$ decay in the NuLeptonSim framework, the angle $\theta_{l}$ is randomly sampled using inverse transform sampling (further detailed in section \ref{sec:stoch}), and the resulting particles in the decay are boosted to the lab frame to calculate the resultant energies. The energy distributions of the charged lepton and the antineutrino produced in a $W^{-}$ decay are shown in Figure \ref{fig:W_decay_distribution}.\\

\begin{figure}
\includegraphics[width=0.98\linewidth]{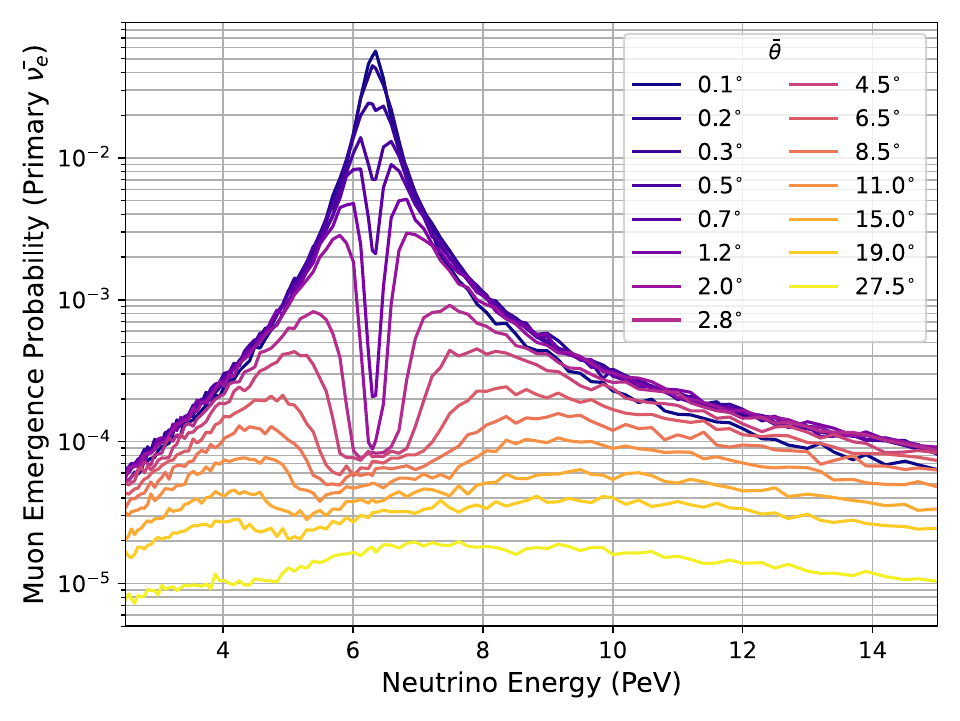}
\includegraphics[width=0.98\linewidth]{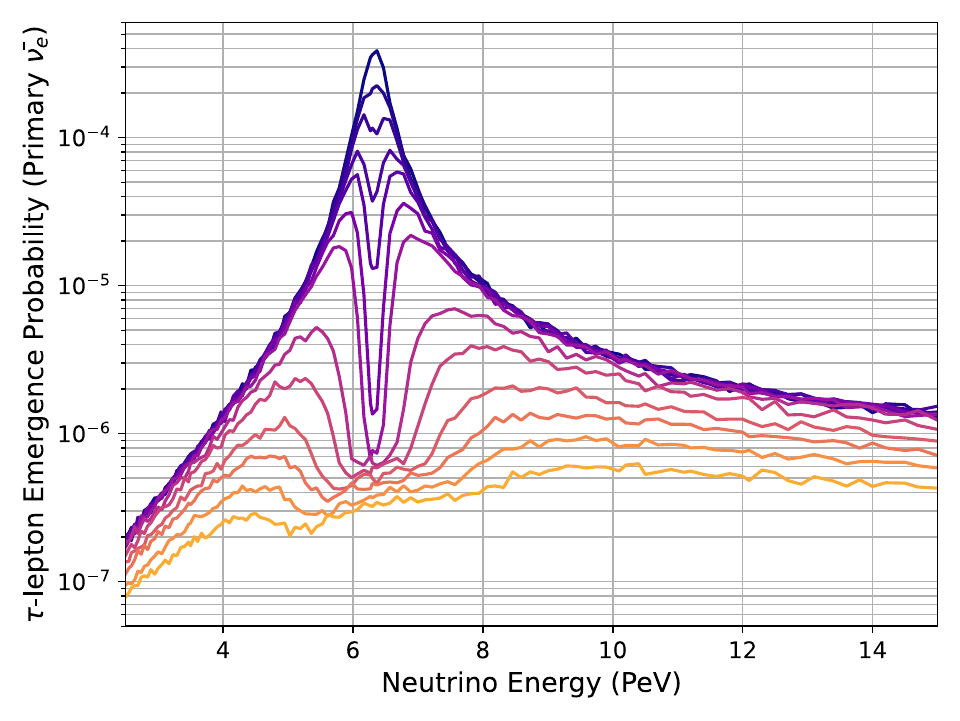}
\includegraphics[width=0.98\linewidth]{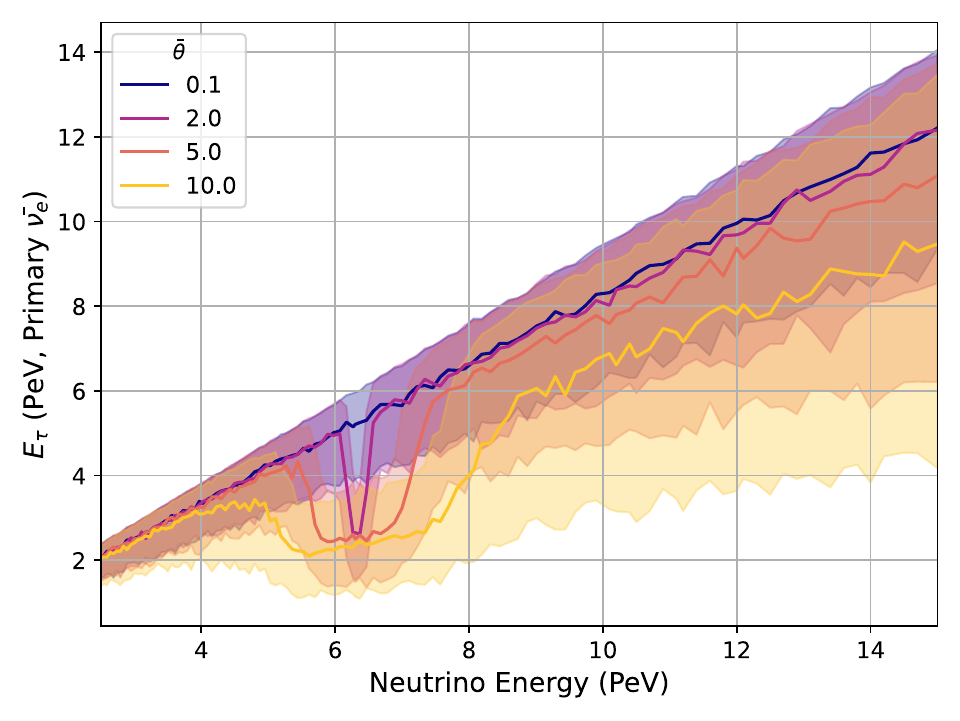}
\caption{Top: Earth emergence probability of muons with energies $E_{\mu}>10$~GeV sourced from primary mono-energetic electron antineutrinos. Middle: Earth emergence probability of $\tau$-leptons with energies $E_{\tau}>100$~TeV sourced from primary mono-energetic electron antineutrinos. Bottom: energy distribution of Earth-emergent $\tau$-leptons, where the solid line represents the mean of the distribution, and the shaded regions represent the $\pm 1\sigma$ deviations. The energy distribution of emergent muons is not shown for reasons explained in the text. The aggregate effect on a given flux is shown in Appendix A.  \label{fig:earth_emergence_GR}}
\end{figure}

We implement the cross section of the Glashow Resonance process into \texttt{NuLeptonSim} using the parameterization given by \cite{Huang_2020, PhysRevD.98.030001}, randomly sample the decay branch of a generated $W$, and determine the energy of the decay products via the process described above, ignoring generated hadrons in future propagation. In Figure \ref{fig:earth_emergence_GR}, we show the Earth emergence probability of muons and $\tau$-leptons given mono-energetic fluxes of primary electron antineutrinos as well as their energy distributions as a function of Earth emergence angle, as simulated with \texttt{NuLeptonSim}.\\

Figure \ref{fig:earth_emergence_GR} shows that, near the resonance energy of 6.3~PeV, there is a relative maxima in Earth emergence probability for both muons and $\tau$-leptons at small Earth emergence angles ($\bar{\theta}<0.3^{\circ}$), and subsequent decreases with increasing $\bar{\theta}$ (deeper trajectories). For an electron antineutrino at the resonance energy, the maximum cross section of the Glashow Resonance process is $\sim 5 \times 10^{-31} \, \mathrm{cm}^{2}$. Given the density of ice ($\rho_{\mathrm{ice}}=0.92 \, \mathrm{g} \, \mathrm{cm}^{-3}$), the average interaction length of the antineutrino is 36~km at this energy, which corresponds to an Earth emergence angle $\bar{\theta} = 0.17^{\circ}$, assuming constant density in the outer ice shell. For electron antineutrinos at the Glashow Resonant energy with trajectories shallower than this angle, the charged leptons produced in the decay of the $W^{-}$ are typically generated near the surface of the Earth and are capable of emerging with minimal energy losses, maximizing the emergence probability. For angles larger than $\sim 0.17^{\circ}$, interactions occur very quickly, and the charged leptons produced in the decay must travel many interaction lengths before emerging, significantly lowering the probability of emergence. The relative maxima on either side of the resonance energy correspond to the ``tuned'' cross sections where the interaction length is comparable to the total grammage provided by the trajectory.\\ 

Comparing Figures \ref{fig:muon_emergence_probability_muons}, \ref{fig:earth_emergence_GR}, and \ref{fig:tau_comp_stoch} (section \ref{sec:stoch}), we note an amplification of emergence probabilities near the resonant energy with respect to emergent muons from muon neutrinos and $\tau$-leptons from tau neutrinos. Specifically, we note that for comparable neutrino energies and small Earth emergence angles, the $\tau$-lepton and muon Earth emergence probability is boosted by a factor of $\sim 30$, in line with our previous approximation. The previous estimation did not consider the fractional energies of the decay products of the $W^{-}$, nor the energy losses of the charged leptons inside the Earth, so the decrease is sensible. Marginal enhancements to the emergence probability are present until Earth emergence angles of $\sim 1^{\circ}$ and between neutrino energies of 5~PeV and 7~PeV.\\

The bottom panel of Figure \ref{fig:earth_emergence_GR} shows that, for emergent $\tau$-leptons sourced via the Glashow Resonance process, the energy distribution is fairly narrow for small Earth emergence angles and is centered around $\sim 0.8 E_{\nu}$. The spread in the energy distribution increases for deeper trajectories, where energy losses feature more prominently, showing significant losses near the resonance energy where the propagation distance of the produced $\tau$-lepton is maximized. We do not show the corresponding energy distribution plot for muons sourced from Glashow Resonance, as the curves for different emergence angles overlap significantly and have a mean energy of 30~TeV across all input antineutrino energies.

\section{Stochastic energy losses} \label{sec:stoch}

\begin{figure}
\includegraphics[width=0.98\linewidth]{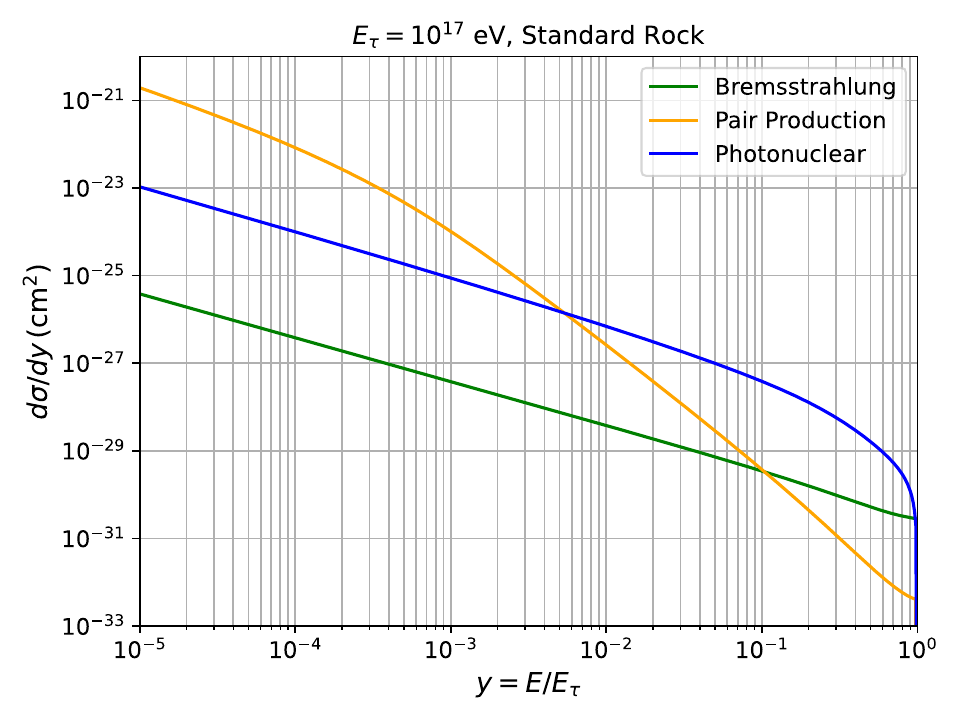}
\includegraphics[width=0.98\linewidth]{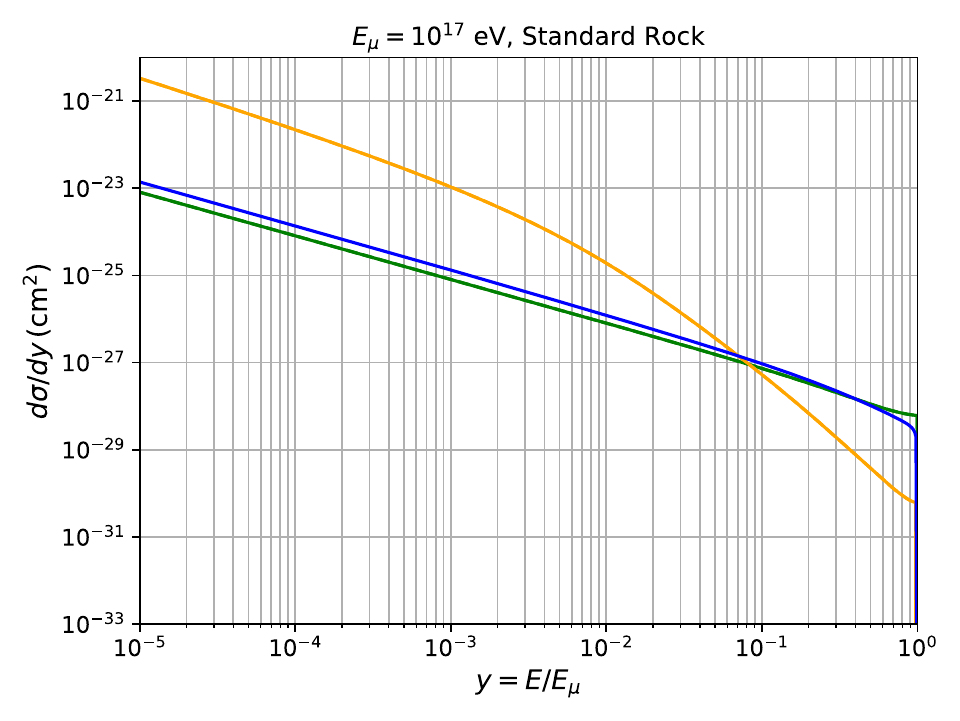}
\caption{Differential cross sections for radiative interactions of $\tau$-leptons (top) and muons (bottom) in standard rock (Z = 11, A = 22), as parameterized in \cite{Petrukhin:1968}, \cite{Kokoulin:1971}, \cite{Abramowicz:1997ms} \label{fig:cross_sections}}
\end{figure}

In the original \texttt{NuTauSim} formalism, the energy losses of propagating $\tau$-leptons within a grammage bin $dX$ were handled via a continuous loss approximation:

\begin{equation}
\Braket{\frac{dE}{dX}} = -\alpha(E)-\beta(E)E
\end{equation}

\noindent where $\alpha$ and $\beta$ represent the respective losses due to ionization and radiative processes. The value of $\alpha$ is roughly a constant $2 \, \mathrm{MeV} \, \mathrm{cm}^{2}/\mathrm{g}$, while $\beta$ is calculated from the differential cross sections of the radiative processes relevant for propagation of charged leptons in dense matter: Bremsstrahlung, $e^{+}/e^{-}$ pair production, and photonuclear interactions. For each type of radiative interaction of either the $\tau$-lepton or muon, the integrated cross section is calculated via:

\begin{equation}
\sigma = \int_{y_{min}}^{y_{max}} \frac{d\sigma}{dy}dy
\label{eq:cross_section}
\end{equation}

\noindent where $\frac{d\sigma}{dy}$ is the differential cross section of the interaction with respect to the inelasticity $y$ (the fractional energy deposited by the lepton $E/E_{l}$), and $y_{min}$ and $y_{max}$ correspond to the physical upper and lower inelasticity limits of the interaction. $\beta$ is calculated via a weighting of the integral in equation \ref{eq:cross_section} by the inelasticity: 

\begin{equation}
\beta(E) = \sum_{i}\beta^{i}(E) = \frac{N}{A} \int_{y_{min}}^{y_{max}} y \frac{d \sigma^{i}(y, E)}{dy} dy
\label{eq:b_calc}
\end{equation}

\noindent where the summation is performed over the three radiative processes. The differential cross sections of these processes for 100~PeV $\tau$-leptons and muons in standard rock (Z = 11, A =22) are shown in Figure \ref{fig:cross_sections}.\\

Under the continuous loss approximation, in a small, fixed step size $dX$, a propagating charged lepton with a fixed energy deposits the same fraction of its energy each step. In the \texttt{NuTauSim} formalism, $dX$ was fixed such that $0.1\%$ of the particle energy was lost in the given step. Moving beyond this approach, it is necessary to model the losses of propagating charged leptons stochastically for two primary reasons:

\begin{itemize}
    \item The continuous loss approximation under-represents the rare instances where large ($>10\%$) energy depositions may occur. By ignoring these interactions, resultant charged leptons have characteristically higher energies compared to those modeled with stochastic losses, resulting in increased propagation ranges and Earth emergence probabilities. Comparisons with \texttt{TauRunner} \cite{Safa:2019ege} and \texttt{NuPyProp} \cite{Patel:2021} show that, for $\tau$-leptons, the effect of using the continuous loss approximation results in a modest ($10\%$) increase of the Earth emergence probability, with an observable impact on outgoing lepton energies. In this updated computation scheme, the propagation of muons is also desired. Muons, on average, experience greater losses than $\tau$-leptons, and will be more strongly influenced by the continuous loss approximation.
    \item As described in section \ref{sec:detectors}, the inclusion of arbitrary trajectories and geometric boundaries allows for tracking neutrinos and charged leptons as they move through a predefined volume. Particles generated in radiative interactions (photons for Bremsstrahlung, electrons and positrons for pair production, and hadrons for photonuclear interactions) can initiate quickly developing subshowers in dense media. Provided the subshowers are of sufficient energy, they may trigger in-ice radio or optical Cherenkov detectors. Modeling this behavior inherently requires stochastic loss modeling.
\end{itemize}

To model the energy losses stochastically, an interaction length $X_{int}$ must be sampled, taking into account all relevant interactions:

\begin{equation}
X_{int} = \frac{A m_{N}}{\sigma_{tot}}
\end{equation}

\noindent where $A$ is the atomic mass number of the material, $m_{N}$ is the nucleon mass, and $\sigma_{tot}$ is the sum of the integrated cross sections of each interaction type, following equation \ref{eq:cross_section}. The interaction type is determined probabilistically by the ratio of the integrated cross sections. Mathematically, this amounts to sampling a random value $r$ between 0 and 1, and selecting the type via:

\begin{align*}
  \mathrm{Type} &=
    \begin{cases}
      \mathrm{Bremsstrahlung } & r\leq\mathlarger{\frac{\sigma_{Brem}}{\sigma_{tot}}}\\
      \mathrm{Pair \, Production } & \mathlarger{\frac{\sigma_{Brem}}{\sigma_{tot}}}<r\leq\mathlarger{\frac{\sigma_{Brem}+\sigma_{PP}}{\sigma_{tot}}}\\
      \mathrm{Photonuclear} & \mathlarger{\frac{\sigma_{Brem}+\sigma_{PP}}{\sigma_{tot}}}<r\leq1
    \end{cases}       
\end{align*}

The inelasticity of the interaction is sampled by calculating the normalized cumulative distribution functions (CDF) of the interaction cross section with respect to inelasticity (following the concept of inverse transform sampling):

\begin{equation}
\mathrm{CDF}(y) = \int_{y_{min}}^{y} \frac{1}{\sigma_{tot}}\frac{d\sigma}{dy}dy
\end{equation}

These CDF values are tabulated for the energy range $10^{11}$~eV to $10^{21}$~eV, spaced in 0.2 decade steps for both $\tau$-leptons and muons. For a given charged lepton energy, a random CDF value is sampled between 0 and 1, and using a 2-Dimensional interpolator, the corresponding inelasticity of the interaction is determined.\\

\begin{figure}
\includegraphics[width=0.98\linewidth]{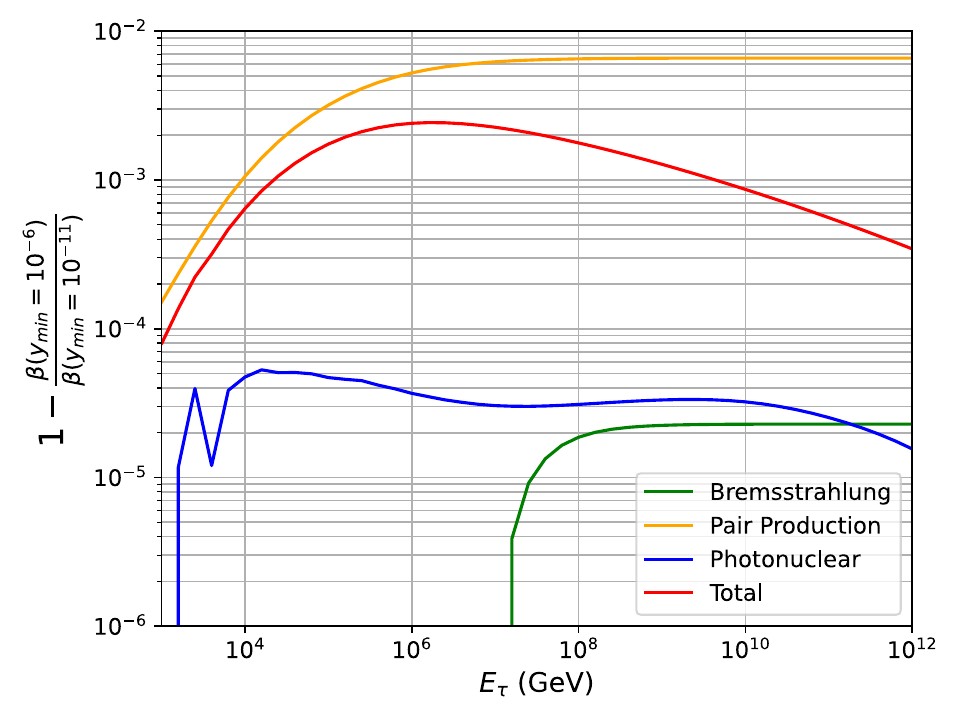}
\includegraphics[width=0.98\linewidth]{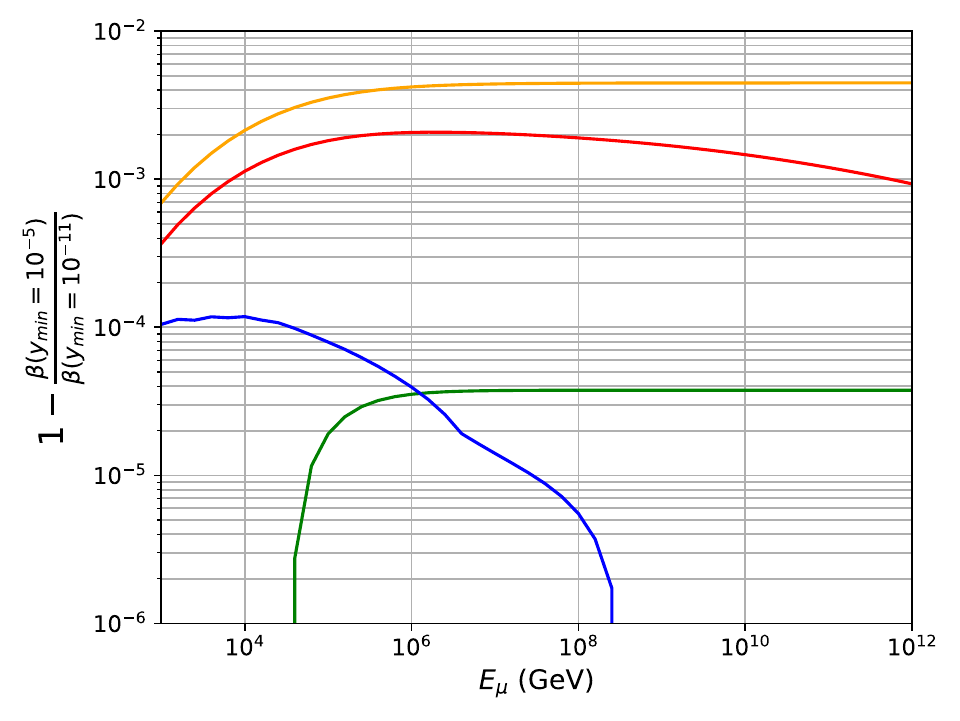}
\caption{Ratio of continuous energy loss factor $\beta(E)$ calculated via equation \ref{eq:b_calc} with $y_{min} = 10^{-6}$ for $\tau$-leptons (top) and $y_{min} = 10^{-5}$ for muons (bottom) in standard rock. \label{fig:losses_ratio}}
\end{figure}

Both the total cross section and the CDF values depend on the value of $y_{min}$, the minimum inelasticity of the radiative process. The value of $y_{min}$ varies for the different interactions and is dependent on the charged lepton mass and energy, and the material properties. The value of $y_{min}$ can grow exceedingly small, particularly for Bremsstrahlung emission, where there is no physical lower limit. Modelling increasingly smaller depositions grows computationally expensive, and has a decreasing effect on the overall propagation. By this logic, we can consider a larger, modified $y_{min}$ such that the behavior of the energy losses is captured accurately while neglecting unimportant losses. We calculate this modified $y_{min}$ to be $10^{-6}$ and $10^{-5}$ for $\tau$-leptons and muons, respectively. Figure \ref{fig:losses_ratio} shows the ratio of $\beta(E)$ values from equation \ref{eq:b_calc} calculated using the modified $y_{min}$ and significantly smaller $y_{min} = 10^{-11}$ (which represents the modeling of all losses to within machine precision) for both $\tau$-leptons and muons in standard rock.\\

\begin{figure}
\includegraphics[width=0.98\linewidth]{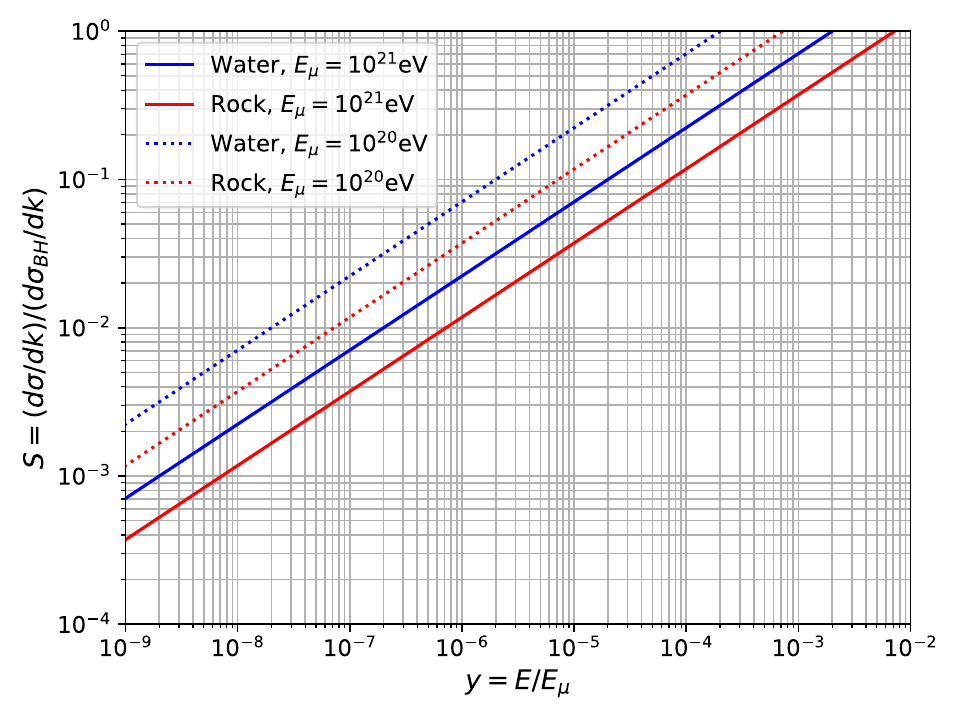}

\caption{Suppression of the muon Bremsstrahlung cross section at high energies due to the LPM effect, compared with the Bethe-Heitler parameterization \cite{Petrukhin:1968}. The LPM suppression is modeled after \cite{Klein_1999, Polityko_2002}. \label{fig:LPM_Suppression}}
\end{figure}

Figure \ref{fig:losses_ratio} quantifies that, by using the increased values for $y_{min}$, the difference in overall energy losses is within $1\%$. For faster computation, CDF tables for $y_{min} = 10^{-5}$ for $\tau$-leptons and $y_{min} = 10^{-4}$ for muons are also included in \texttt{NuLeptonSim}, and result in energy loss differences within $\sim 3\%$. For propagation inside detection volumes where expected energy thresholds are low (such as optical Cherenkov experiments) or where strong accuracy is desired, CDF tables for $y_{min}=10^{-7}$ and $y_{min}= 10^{-9}$ are also included for both $\tau$-leptons and muons. When very small $y$ values are considered, the (Landau–Pomeranchuk–Migdal) LPM effect for muons is non-negligible, particularly at high energies, and is thus included in the corresponding tables within \texttt{NuLeptonSim} \cite{Klein_1999, Polityko_2002}. Figure \ref{fig:LPM_Suppression} shows the Bremsstrahlung suppression factor due to the LPM effect for muons propagating through water and rock at high energies. The energy threshold for LPM suppression of pair production by muons is significantly higher than that for Bremsstrahlung emission, and is negligible in most cases, but is included in \texttt{NuLeptonSim} for completeness. LPM suppression of $\tau$-lepton radiative loss processes are ignored due to the extremely high energy thresholds required. \texttt{NuLeptonSim} is also configured to optionally use the continuous loss approximation where desired.\\ 

\begin{figure}
\includegraphics[width=0.98\linewidth]{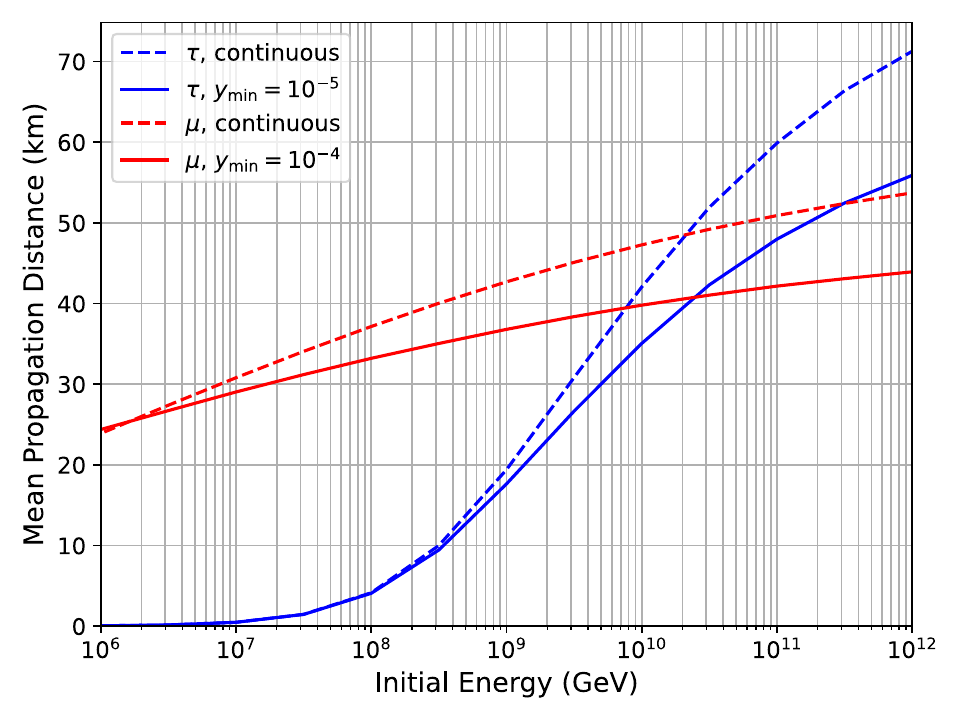}

\caption{Mean range (propagation distance until either decay or energy below a threshold of $10^2$ GeV) for $\tau$-leptons (blue) and muons (red) in an infinite plane of ice with density 0.92 $\mathrm{g}/\mathrm{cm}^3$ calculated using continuous and stochastic modelling of energy losses. \label{fig:sto_vs_cont}}
\end{figure}

To demonstrate the effect of charged lepton energy loss modeling, Figure \ref{fig:sto_vs_cont} shows the mean in-ice propagation range of $\tau$-leptons and muons using both the continuous loss approximation and the stochastic modeling. As expected, the propagation range is increased when using the continuous loss approximation, with the discrepancy between the two methods increasing with energy.\\

\begin{figure}
\includegraphics[width=0.98\linewidth]{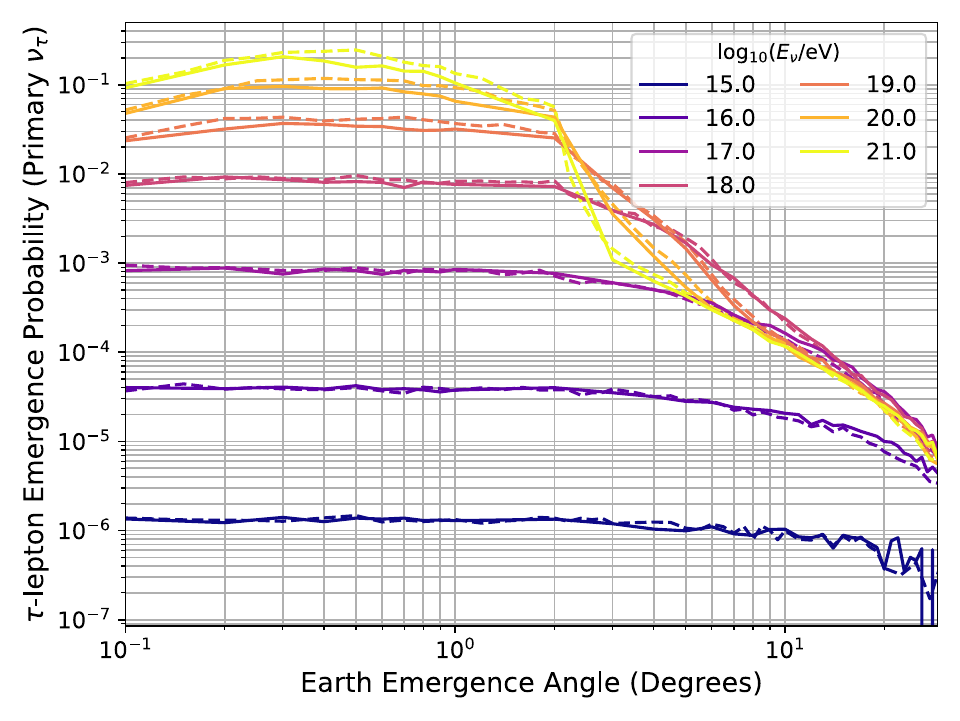}
\includegraphics[width=0.98\linewidth]{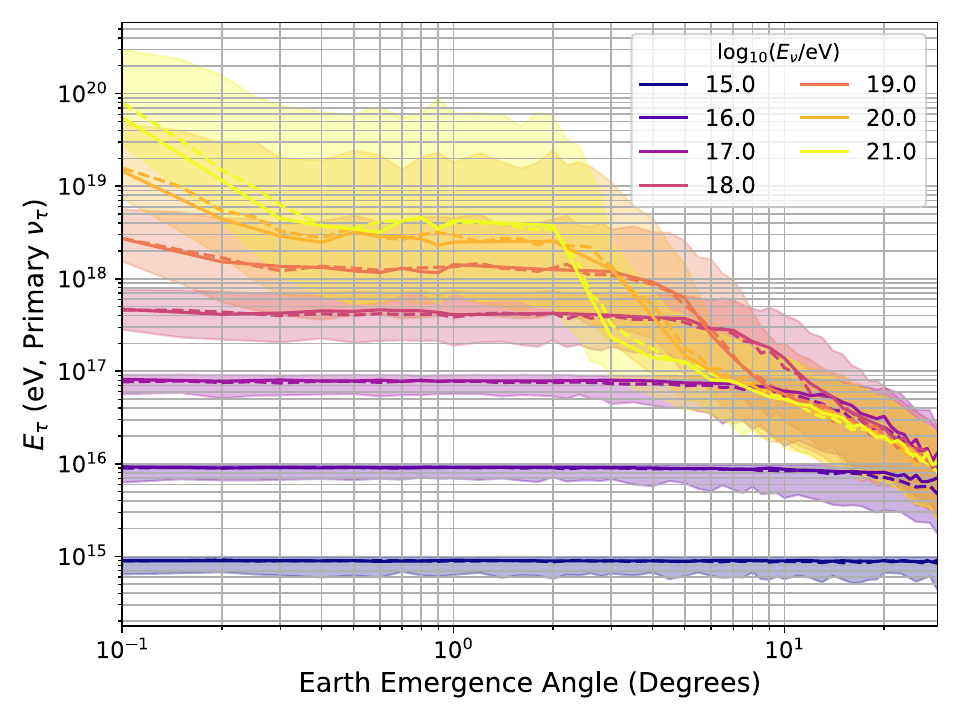}
\caption{Top: Earth emergence probability of $\tau$-leptons with energies $E_{\tau}>100$~TeV sourced from primary mono-energetic tau neutrinos using both continuous (dashed line) and stochastic (solid line) energy loss modeling. Bottom: energy distribution of Earth-emergent $\tau$-leptons, where the central line represents the mean of the distribution, and the shaded regions represent the $\pm 1\sigma$ deviations. The aggregate effect on a given flux is shown in Appendix A. \label{fig:tau_comp_stoch}}
\end{figure}

In Figure \ref{fig:tau_comp_stoch}, we show the Earth emergence probability of $\tau$-leptons given mono-energetic fluxes of primary tau neutrinos as well as their energy distributions as a function of Earth emergence angle, as simulated with \texttt{NuLeptonSim} using both continuous and stochastic energy loss modeling.\\

Figure \ref{fig:tau_comp_stoch} shows that using the continuous energy loss approximation overestimates $\tau$-lepton emergence probabilities for primary tau neutrino energies above 100~PeV, reaching up to $15\%$ discrepancies at $E_{\nu} = 10^{21}$~eV and small ($\bar{\theta} < 2^{\circ}$) Earth emergence angles, due to the decreased propagation ranges of the $\tau$-lepton under stochastic energy loss modeling. Similarly, the energy distribution of Earth emergent $\tau$-leptons shifts to lower energies under stochastic energy loss modeling, being able to properly account for catastrophic losses. However, using the continuous loss approximation produces accurate results over a wide parameter space, and is sufficient for the majority of use cases. This is in agreement with the results of \texttt{NuPyProp}, and suggests that for deeper emergence angles, where extensive simulation is required to produce high statistics, computations can be made less costly by using the continuous loss approximation \cite{Garg_2023}.\\

Sections \ref{sec:mu_from_mu} and \ref{sec:mu_from_tau} suggest that the flux of emergent muons begins to dominate over $\tau$-leptons below energies of 100~PeV, where the effect of stochastic losses is minimal. Thus, the effect of stochastic energy losses on emergent muons was not calculated in this work and Figures \ref{fig:muon_emergence_probability_muons}, \ref{fig:muon_emergence_probability_taus}, and \ref{fig:earth_emergence_GR} were generated using the continuous loss approximation to reduce computation time.

\section{Benchmarking Results}

\texttt{NuLeptonSim} is fast in comparison to other propagation codes, particularly when using the continuous energy loss approximation for propagating leptons. Table \ref{tab:runtimes} shows the benchmarked simulation times for high statistic scans across relevant Earth emergence angles on a personal machine, computed for tau neutrinos and muon neutrinos, modeling energy losses continuously and through the stochastic approach defined above. Regarding the range of Earth emergence angles simulated, we specifically choose the same range as that benchmarked for the \texttt{NuPyProp} computation scheme: 1$^{\circ}$, 2$^{\circ}$, 3$^{\circ}$, 5$^{\circ}$, 7$^{\circ}$, 10$^{\circ}$, 12$^{\circ}$, 15$^{\circ}$, 17$^{\circ}$, 20$^{\circ}$, 25$^{\circ}$, 30$^{\circ}$, 35$^{\circ}$ \cite{Garg_2023}.\\

\begin{table}[ht]
\begin{center}
\begin{tabular}{ |l|l|l|l| }
\hline
\multicolumn{4}{|l|}{Earth Emergence Angles: $1^{\circ}-35^{\circ}$} \\
\multicolumn{4}{|l|}{Statistics: $10^{8}$ primary neutrinos} \\
\multicolumn{4}{|l|}{CPU: Ryzen 5 5600X, 6 cores 12 threads} \\
\multicolumn{4}{|l|}{$^{*}E_{\mu}>$ 10~GeV, 1~TeV for \texttt{NuLeptonSim}, \texttt{NuPyProp} \cite{Garg_2023}} \\
 \hline
 \textbf{Energy}, $\boldsymbol{P_{\mathrm{max}}}$ & \textbf{Neutrino Type} & \textbf{\texttt{NLS} Time} & \textbf{\texttt{NPP} Time} \\
 \hline
 $10^{7}$~GeV & $\nu_{\tau}$ (continuous) & 0.183~h & 0.88~h\\ 
 \cline{1-1}
  $P_{\texttt{NLS}} \sim 4 \times 10^{-5}$ & $\nu_{\tau}$ ($y_{min} = 10^{-6}$) & 4.36~h & 1.07~h\\
 $P_{\texttt{NPP}} \sim 4 \times 10^{-5}$ & & & \\
 \hline
  $10^{8}$~GeV & $\nu_{\tau}$ (continuous) & 0.416~h & 5.51~h\\
   \cline{1-1}
  $P_{\texttt{NLS}} \sim 9 \times 10^{-4}$ & $\nu_{\tau}$ ($y_{min} = 10^{-6}$) & 31.83~h & 6.18~h\\
 $P_{\texttt{NPP}} \sim 9 \times 10^{-4}$ & & & \\
 \hline
   $10^{9}$~GeV & $\nu_{\tau}$ (continuous) & 2.55~h & 19.11~h\\
    \cline{1-1}
  $P_{\texttt{NLS}} \sim 9 \times 10^{-3}$ & $\nu_{\tau}$ ($y_{min} = 10^{-6}$) & \longdash[8] & 27.96~h\\ 
   $P_{\texttt{NPP}} \sim 9 \times 10^{-3}$ & & & \\
 \hline
    $10^{10}$~GeV & $\nu_{\tau}$ (continuous) & 5.35~h & 35.59~h\\
     \cline{1-1}
  $P_{\texttt{NLS}} \sim 3.5 \times 10^{-2}$ & $\nu_{\tau}$ ($y_{min} = 10^{-6}$) & \longdash[8] & 49.80~h\\ 
   $P_{\texttt{NPP}} \sim 3.5 \times 10^{-2}$ & & & \\
 \hline
 \hline
 $10^{7}$~GeV & $\nu_{\mu}$ (continuous) & 0.192~h & 3.19~h\\
  \cline{1-1}
  $P_{\texttt{NLS}}^{*} \sim 3 \times 10^{-3}$ & $\nu_{\mu}$ ($y_{min} = 10^{-4}$) & 4.36~h & \longdash[8] \\ 
   $P_{\texttt{NPP}}^{*} \sim 2.5 \times 10^{-3}$ & & & \\
 \hline
  $10^{8}$~GeV & $\nu_{\mu}$ (continuous) & 0.75~h & 5.17~h\\
   \cline{1-1}
  $P_{\texttt{NLS}}^{*} \sim 9 \times 10^{-3}$ & $\nu_{\mu}$ ($y_{min} = 10^{-4}$) & 29.43~h & \longdash[8]\\ 
   $P_{\texttt{NPP}}^{*} \sim 8 \times 10^{-3}$ & & & \\
 \hline
   $10^{9}$~GeV & $\nu_{\mu}$ (continuous) & 2.55~h & 7.42~h\\
    \cline{1-1}
  $P_{\texttt{NLS}}^{*} \sim 2.5 \times 10^{-2}$ & $\nu_{\mu}$ ($y_{min} = 10^{-4}$) & \longdash[8] & 111.77~h \\ 
   $P_{\mathrm{NPP}}^{*} \sim 2 \times 10^{-2}$ & & & \\
 \hline
    $10^{10}$~GeV & $\nu_{\mu}$ (continuous) & 5.38~h & 9.76~h\\
     \cline{1-1}
  $P_{\texttt{NLS}}^{*} \sim 5 \times 10^{-2}$ & $\nu_{\mu}$ ($y_{min} = 10^{-4}$) &  \longdash[8] & 98.17~h\\ 
   $P_{\texttt{NPP}}^{*} \sim 4 \times 10^{-2}$ & & & \\
 \hline
 
\end{tabular}
\end{center}
\caption{Benchmarked runtimes and max Earth emergence probabilities for primary tau neutrinos and muon neutrinos simulated with \texttt{NuLeptonSim (NLS)}. Runtimes are shown for modeling energy losses continuously and through the stochastic approach defined above. Comparisons to the \texttt{NuPyProp (NPP)} computation scheme are also provided. The muon energy threshold for continued propagation varies between the schemes: 10~GeV is used in this work while 1~TeV is used for \texttt{NuPyProp}. This variation leads to a $\sim 15\%$ increase in muon Earth emergence probabilities with respect to \texttt{NuPyProp}.}
\label{tab:runtimes}
\end{table}

Table \ref{tab:runtimes} shows that modeling energy losses stochastically reduces the computation performance of \texttt{NuLeptonSim} by over an order of magnitude in some cases, but remains competitive with other propagation codes (in this case, comparing to the \texttt{NuPyProp} computation scheme). Using the continuous energy loss approximation, \texttt{NuLeptonSim} is faster than \texttt{NuPyProp} by roughly an order of magnitude, but can result in deviations of the Earth emergence probability within $5-15\%$ at the highest energies with respect to the more physical stochastic modeling of energy losses.\\

\section{Modeling in-ice detector geometries} \label{sec:detectors}
The \texttt{NuTauSim} formalism was originally developed to facilitate propagation of neutrinos as it pertains to high-altitude (mountain-top, balloon-based, space-based) observation points that aim to indirectly measure EAS sourced from Earth-emergent particles.  In this scenario, the observation method is blind to interactions inside the Earth, and thereby only requires the Earth emergence probability and energy distributions of emergent particles to model sensitivity. \\

For in-ice experiments, such as ARA \cite{ARA}, IceCube \cite{Aartsen:2013jdh}, RNO-G \cite{RNOG}, and RET \cite{RET} and high-altitude experiments sensitive to in-ice Askaryan emission (such as PUEO \cite{PUEO}), interactions inside the Earth are critical to determining neutrino sensitivity. In the \texttt{NuLeptonSim} framework, it is possible to record in-ice energy depositions from neutrino interactions and charged lepton interaction processes, inherently requiring the use of the stochastic energy loss modeling detailed in section \ref{sec:stoch}. In \texttt{NuTauSim}, the geometry of a trajectory was defined uniquely by the Earth emergence angle $\bar{\theta}$. Because interactions can occur anywhere in the ice shell, this geometry is updated in the \texttt{NuLeptonSim} framework to allow for the selection of a trajectory based on start and end positions anywhere in or on the Earth, with the step size $\overrightarrow{dl}$ defined to be 3-Dimensional. Using 3-Dimensional trajectories, it is possible to define geometric boundaries, and only record those in-ice interactions which occur inside the volume, limiting studies to events which have sensible chances of observation. In \texttt{NuLeptonSim}, example 3-D geometries are predefined that can be used in these sensitivity studies including: cylinders, spheres, and rectangular prisms. Interactions that occur inside the volume can be saved and output to simulations which model the signals produced by the subsequent in-ice subshowers for detector sensitivity studies.\\

\texttt{NuLeptonSim} has been used to calculate the ARA-5 sensitivity to all flavors of primary neutrinos using the procedures above, taking into consideration events induced by radiative losses of propagating charged leptons \cite{Cummings:ARA, AbbyICRC}. This study demonstrated sensible agreement with scaled estimates of the A2 (single ARA station) neutrino sensitivity, calculated using only primary neutrino interaction vertices \cite{ARA_EA_1}, and suggested a factor of $\sim 2$ improvement in effective area when considering events triggered by charged lepton interactions at high energies ($E_{\nu} > 10^{19}$~eV), in line with other work regarding secondary interactions in ice \cite{secondaries}. Using the same dataset generated by \texttt{NuLeptonSim}, further studies regarding the topology of observed events in the ARA-5 detector are currently ongoing, and will qualify the improvements gained from observing multi-deposition candidates.

\section{Discussion and Conclusions}
The \texttt{NuLeptonSim} framework builds on the foundation of the successful \texttt{NuTauSim} computation scheme to provide a fast and robust simulation pipeline for neutrino and charged lepton propagation in the Earth, covering a wide range of use cases and producing compatible results with the vast array of existing propagating codes \cite{Garg_2023}. In this work, we have highlighted the principal updates that constitute the \texttt{NuLeptonSim} framework, including the modeling of new particles and interaction processes, multi-particle tracking, stochastic energy loss modeling, and 3-dimensional tracking. We have detailed the use cases of each of these inclusions and how they are implemented in the overall propagation scheme.\\

The inclusion of muon propagation in \texttt{NuLeptonSim} provides another channel by which to model observation of high energy neutrinos, either via interactions in ice, or Earth-emergence and interaction in the atmosphere. By implementing muon propagation in \texttt{NuLeptonSim}, we demonstrate that neutrino-sourced muons represent a non-negligible source of potentially measurable events, with Earth-emergence probabilities outpacing those of $\tau$-leptons up to neutrino energies of 100~PeV, and remaining comparable for higher energies. Despite the increased energy losses of muons in the Earth with respect to $\tau$-leptons, Earth emergent (and in-ice) muons can often retain a significant amount of the parent neutrino energy, giving them the chance to trigger a given experiment. While this impact has previously been quantified for orbital and sub-orbital optical Cherenkov experiments \cite{Cummings:2021}, this work demonstrates the potential for muon measurements from mountain-based and in-ice experiments using both the optical Cherenkov and radio approaches. In addition, we show here that the muon flux from primary tau neutrinos is non-negligible, with muon emergence probabilities being larger than $\tau$-leptons for neutrino energies below 100~PeV for most observation angles. The additional channels of muon initiated EAS sourced from Earth-skimming neutrinos will produce a flavor impurity that should be considered in evaluating detection capabilities for a wide range of developing experiments.\\

The Glashow Resonance effect is also included in \texttt{NuLeptonSim}, and for small Earth emergence angles, provides an amplification of the Earth emergence probability for both $\tau$-leptons and muons of $\sim 30$ at the resonant anti-neutrino energy of 6.3~PeV with respect to the standard charged current interactions. Assuming a constant density of ice, we calculate that the critical angle where the interaction length of Glashow Resonance at the resonant energy matches the path length through the Earth to be $\bar{\theta} = 0.17^{\circ}$. For angles significantly greater than this, interactions occur very quickly, leading to large path lengths through the Earth and the domination of charged lepton energy losses, resulting in local minima in Earth emergence probabilities. By increasing or decreasing the anti-neutrino energy from resonance, the cross section decreases, and mirrored local maxima appear in the Earth emergence probability. $\tau$-leptons and muons produced by Glashow Resonance are expected to improve the neutrino sensitivity of mountain-top and sub-orbital optical Cherenkov experiments up to $1^{\circ}$ Earth emergence angles within the energy range 5-7~PeV.\\

We have included the ability to model charged lepton energy losses using a stochastic approach in \texttt{NuLeptonSim}, which better accounts for rare, high energy depositions that are ignored under the continuous loss approximation. In this approach, we consider Bremmstrahlung, $e^{\pm}$ pair production, and photonuclear processes separately, and randomly sample energy depositions based on the differential cross sections with respect to the inelasticity of the interaction. We demonstrate that, for a given interaction, lower inelasticity bounds can be used in place of physical lower bounds while disregarding $<1\%$ of total energy losses to improve computational performance. The mean propagation range of charged leptons is decreased when modeling their losses stochastically, particularly so at high energies, which results in Earth emergence probabilities that are systematically lower by $\sim5-15\%$ at the highest energies with respect to those calculated using the continuous loss approximation. Combined with the 3-dimensional propagation modeling included in \texttt{NuLeptonSim}, the stochastic modeling of energy losses allows for tracking all interactions inside predefined volumes within the Earth, which has been used to help evaluate the sensitivity of in-ice experiments to primary neutrino interactions as well as secondary interactions from propagating charged leptons.\\


\texttt{NuLeptonSim} is a complementary framework to existing propagation codes, and has been used to provide probability and energy lookup tables for various simulations, including the \texttt{tapioca} radio Monte Carlo simulation \cite{tapioca} and recently, the \texttt{NuSpaceSim} comprehensive neutrino simulation package \cite{krizmanic}. In addition, \texttt{NuLeptonSim} has been used in conjunction with \texttt{PyREx} to calculate the neutrino acceptance of the ARA-5 detector. \texttt{NuLeptonSim} takes additional steps to more accurately model the processes present during neutrino propagation in the Earth in order to improve neutrino sensitivity estimates of current and planned experiments. The robust simulation pipeline makes it easier to consider arbitrary experimental designs, and the inclusion of all relevant production channels of charged leptons from all flavors of neutrinos provides the largest bandwidth for discovery potential. 

\begin{acknowledgments}
S.~Wissel, A.~Cummings, and R.~Krebs acknowledge funding support through NSF Awards \#2033500, \#2111232 and NASA Awards \#80NSSC20K0925 and \#80NSSC22K1519. J.~Alvarez-Mu\~{n}iz and E.~Zas have received financial support from Mar\'\i a de Maeztu grant CEX2023-001318-M funded by MICIU/AEI /10.13039/501100011033; Xunta de Galicia, Spain (CIGUS Network of Research Centers and Consolidaci\'on 2021 GRC GI-2033 ED431C-2021/22 and 2022 ED431F-2022/15); Feder Funds; Ministerio de Ciencia, Innovaci\'on y Universidades/Agencia Estatal de Investigaci\'on, Spain (PID2022-140510NB-I00, PCI2023-145952-2); and European Union ERDF. W. Carvalho Jr. would like to acknowledge Polish National Agency for Academic Exchange within Polish Returns Program no. PPN/PPO/2020/1/00024/U/00001,174; National Science Centre Poland for NCN OPUS grant no. 2022/45/B/ST2/0288.
\end{acknowledgments}

\section*{Appendix A: Exiting Lepton Fluxes}

We have, so far, only considered fluxes of mono-energetic neutrinos when considering properties of Earth-emergent leptons. As in \cite{Alvarez:2018, Alvarez:2019}, we find it useful to calculate fluxes of leptons from a given flux of neutrinos to demonstrate aggregate behavior. In principle, for a given neutrino flux $\phi_{\nu}$ at an Earth emergence angle $\bar{\theta}$, the flux of charged leptons $\phi_{\ell}$ be calculated via:

\begin{equation}
\phi_{\ell}(E_{\ell}, \bar{\theta}) = \phi_{\nu}(E_{\nu}) P(E_{\nu}, \bar{\theta}) \frac{\mathrm{d}E_{\nu}}{\mathrm{d}E_{\ell}}(E_{\nu}, \bar{\theta})
\end{equation}

where $P$ is the Earth emergence probability and $\frac{\mathrm{d}E_{\nu}}{\mathrm{d}E_{\ell}}$ is the charged lepton energy distribution. For these parameters, we use those calculated in the previous sections of this work. For the input flux of neutrinos, we consider the best-fit cosmogenic flux calculated by the Pierre-Auger collaboration, assuming a source evolution of $m = +3$, and a maximum redshift range of $z_{\mathrm{max}} \in [1, 5]$ \cite{Abdul_Halim_2023}. We assume here a 1:1:1 neutrino flavor ratio at Earth, and even amounts of neutrinos and anti-neutrinos.\\

\begin{figure}
\includegraphics[width=0.98\linewidth]{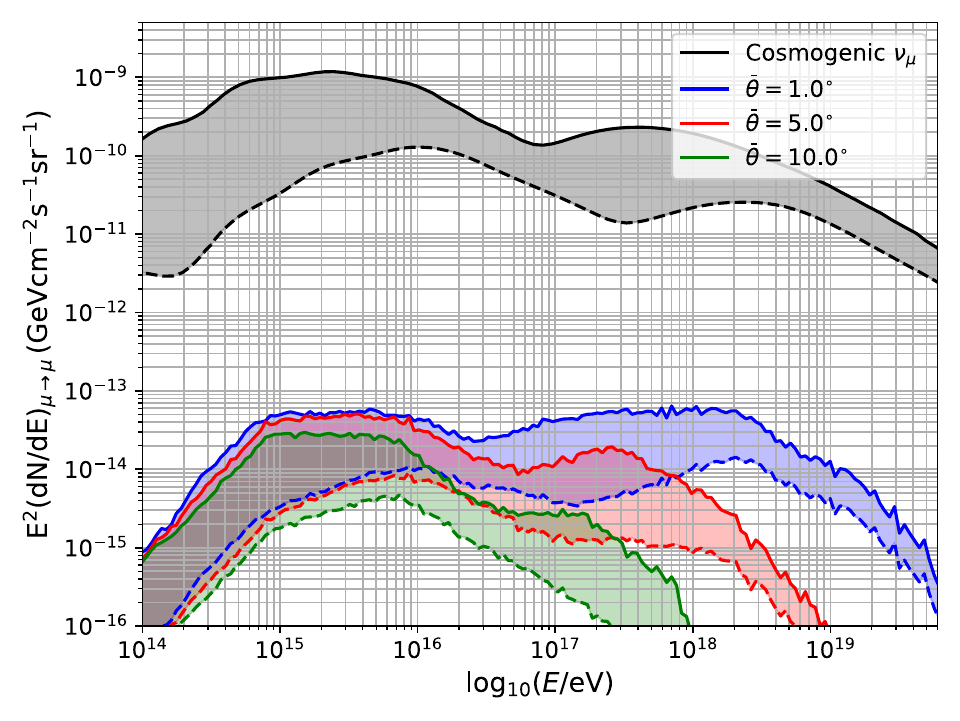}
\caption{Earth-emergent muon flux, calculated in the range $\bar{\theta} \in [1^{\circ}, 5^{\circ}, 10^{\circ}]$, given the Auger best-fit flux of cosmogenic muon neutrinos, assuming a source evolution of $m = +3$, and a maximum redshift range of $z_{\mathrm{max}} \in [1, 5]$ \cite{Abdul_Halim_2023}.  \label{fig:cosmogenic_mu_mu}}
\end{figure}

\begin{figure}
\includegraphics[width=0.98\linewidth]{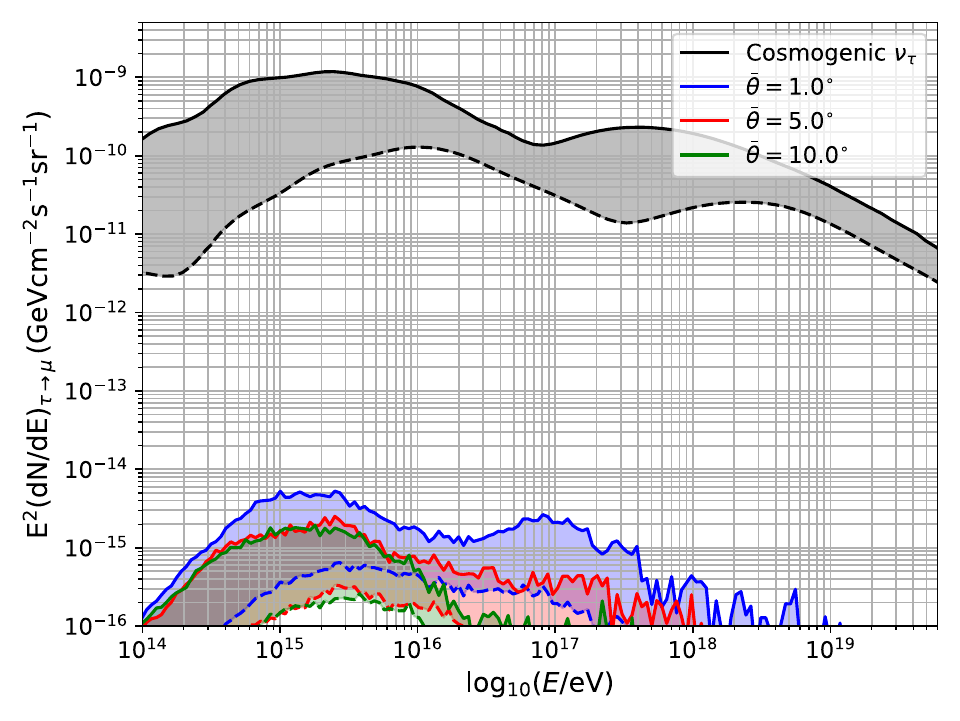}
\caption{Earth-emergent muon flux, calculated in the range $\bar{\theta} \in [1^{\circ}, 5^{\circ}, 10^{\circ}]$, given the Auger best-fit flux of cosmogenic tau neutrinos, assuming a source evolution of $m = +3$, and a maximum redshift range of $z_{\mathrm{max}} \in [1, 5]$ \cite{Abdul_Halim_2023}. \label{fig:cosmogenic_tau_mu}}
\end{figure}

\begin{figure}
\includegraphics[width=0.98\linewidth]{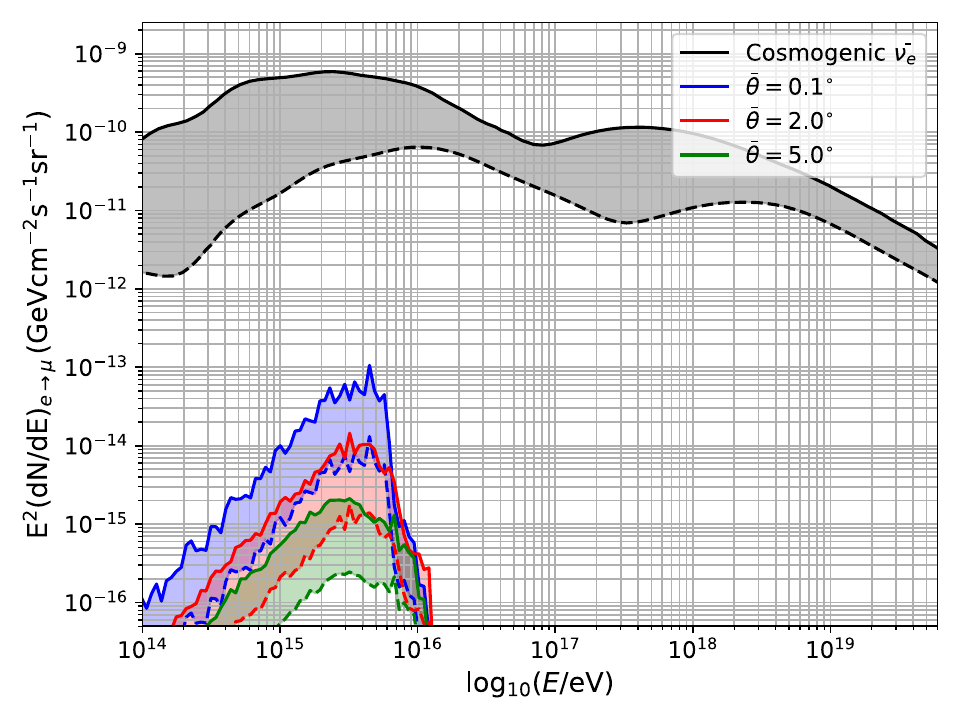}
\caption{Earth-emergent muon flux, calculated in the range $\bar{\theta} \in [0.1^{\circ}, 2^{\circ}, 5^{\circ}]$, given the Auger best-fit flux of cosmogenic electron anti-neutrinos, assuming a source evolution of $m = +3$, and a maximum redshift range of $z_{\mathrm{max}} \in [1, 5]$ \cite{Abdul_Halim_2023}. \label{fig:cosmogenic_e_mu}}
\end{figure}

\begin{figure}
\includegraphics[width=0.98\linewidth]{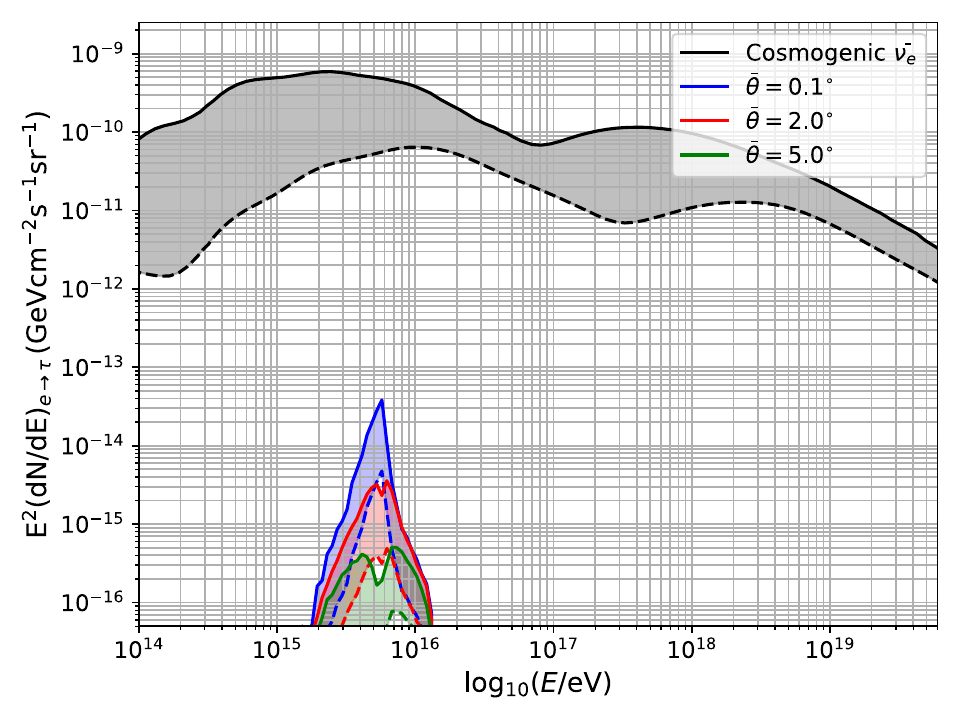}
\caption{Earth-emergent $\tau$-lepton flux, calculated in the range $\bar{\theta} \in [0.1^{\circ}, 2^{\circ}, 5^{\circ}]$, given the Auger best-fit flux of cosmogenic electron anti-neutrinos, assuming a source evolution of $m = +3$, and a maximum redshift range of $z_{\mathrm{max}} \in [1, 5]$ \cite{Abdul_Halim_2023}. \label{fig:cosmogenic_e_tau}}
\end{figure}

\begin{figure}
\includegraphics[width=0.98\linewidth]{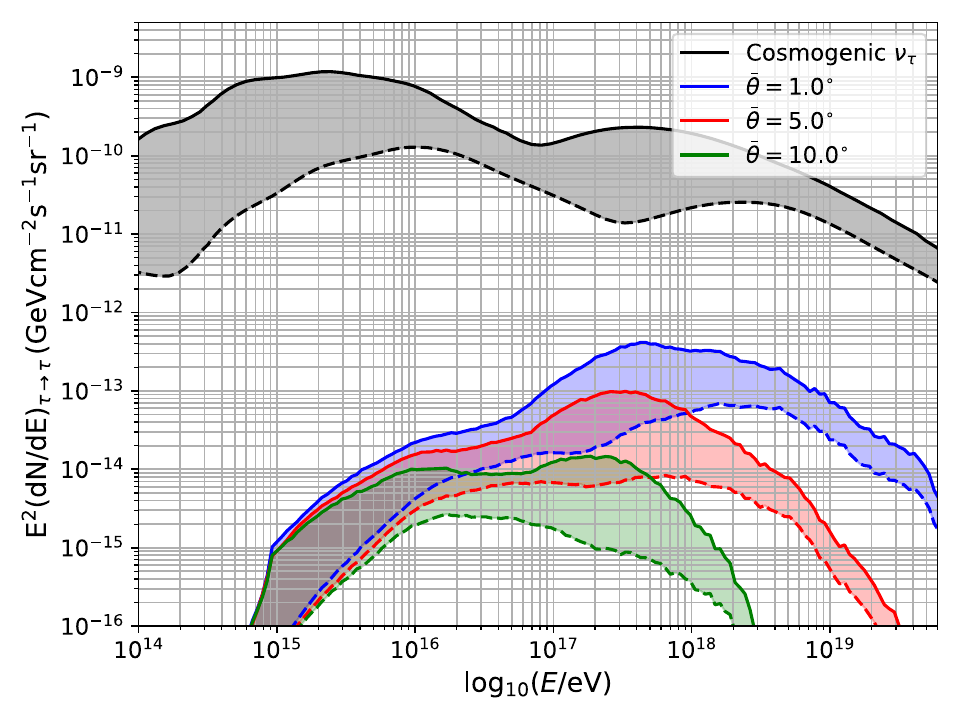}
\caption{Earth-emergent $\tau$-lepton flux, calculated in the range $\bar{\theta} \in [1^{\circ}, 5^{\circ}, 10^{\circ}]$, given the Auger best-fit flux of cosmogenic tau neutrinos, assuming a source evolution of $m = +3$, and a maximum redshift range of $z_{\mathrm{max}} \in [1, 5]$ \cite{Abdul_Halim_2023}. \label{fig:cosmogenic_tau_tau}}
\end{figure}

The Earth-emergent fluxes of charged leptons given this input neutrino flux are calculated for Earth emergence angles $\bar{\theta} \in [1^{\circ}, 5^{\circ}, 10^{\circ}]$ for the channels $\nu_{\mu} \rightarrow \mu$ (Figure \ref{fig:cosmogenic_mu_mu}), $\nu_{\tau} \rightarrow \mu$ (Figure \ref{fig:cosmogenic_tau_mu}), and $\nu_{\tau} \rightarrow \tau$ (Figure \ref{fig:cosmogenic_tau_tau}) and $\bar{\theta} \in [0.1^{\circ}, 2^{\circ}, 5^{\circ}]$ for the Glashow Resonance channels $\bar{\nu_{e}} \rightarrow \mu$ (Figure \ref{fig:cosmogenic_e_mu}) and $\bar{\nu_{e}} \rightarrow \tau$ (Figure \ref{fig:cosmogenic_e_tau}). For the Glashow Resonance channels, we consider only the flux of anti-neutrinos, while other channels treat neutrinos and anti-neutrinos identically. It is also noted here that the results of the $\nu_{\tau} \rightarrow \tau$ channel should not be directly compared to the results presented in \cite{Alvarez:2018, Alvarez:2019}, as the energy distribution of Earth-emergent $\tau$-leptons was not handled properly in this previous work.\\

Figures \ref{fig:cosmogenic_mu_mu}-\ref{fig:cosmogenic_tau_tau} exhibit many of the conclusions reached by analyzing the mono-energetic calculations performed in the rest of this work. These include: i) the flux of Earth-emergent muons resulting from muon neutrino interactions is comparable to that of $\tau$-leptons sourced from tau neutrinos at high energies and amplified below 100~PeV energies ii) the flux of muons sourced from tau neutrino interactions is non-negligible and greater than the flux of $\tau$-leptons sourced from tau neutrinos below a few PeV iii) within a narrow band in neutrino energy, the Glashow resonance process substantially contributes to the flux of Earth-emergent charged leptons and is therefore a non-negligible interaction process for instruments capable of detecting $\mathcal{O}$(PeV) signals.

\clearpage
\newpage
\bibliography{references.bib}

\end{document}